\documentclass[11pt,groupcitations]{article}
\RequirePackage{latexsym,amsmath,amssymb,cite}
\RequirePackage[dvipsnames,usenames]{color}
\usepackage{a4wide}
\usepackage{amssymb}
\usepackage{bm}
\usepackage{dcolumn}
\usepackage{amsfonts}
\usepackage{graphicx,epsfig}
\usepackage{psfrag}
\usepackage{multicol}
\usepackage{tikz}
\usetikzlibrary{arrows,shapes}
\usetikzlibrary{matrix,arrows}
\newcommand{\be}{\begin{eqnarray}}
\newcommand{\ee}{\end{eqnarray}}
\newcommand{\bra}[1]{\mbox{$\langle\, #1 \mid$}}

\newcommand{\ket}[1]{\mbox{$\mid #1\,\rangle$}}

\newcommand{\pro}[2]{\mbox{$\langle\, #1 \mid #2\,\rangle$}}
\newcommand{\expec}[1]{\mbox{$\langle\, #1\,\rangle$}}

\renewcommand{\d}{\mbox{${\rm d}$}} %d differenziale non corsivo in math mode
\newcommand{\lp}{\ell_{\rm p}}
\newcommand{\mpl}{m_{\rm p}}
\newcommand{\gn}{G_{\rm N}}
\newcommand{\rh}{R_{\rm H}}

\DeclareMathOperator\Erf{Erf}
\title{\bf Horizon Quantum Mechanics of collapsing shells}
\author{Roberto~Casadio$^a$\thanks{E-mail: casadio@bo.infn.it}$\ $
and
Octavian Micu$^b$\thanks{E-mail: octavian.micu@spacescience.ro}
\\
\\
{\em $^a$Dipartimento di Fisica, Universit\`a di
Bologna and I.N.F.N., Sezione di Bologna,}
\\
{\em via Irnerio~46, I-40126 Bologna, Italy}
\\
\\
{\em $^b$Institute of Space Science, Bucharest, Romania}
\\
{\em P.O. Box MG-23, RO-077125 Bucharest-Magurele, Romania}}
\begin{document}
\maketitle
\begin{abstract}
We study the probability that a horizon appears when concentric shells of matter collide,
by computing the horizon wave-function of the system.
We mostly consider the collision of two ultra-relativistic shells, both shrinking and expanding,
at the moment their radii are equal, and find a probability that the system is a black hole
which is in qualitative agreement with what one would expect according to the hoop conjecture
and the uncertainty principle of quantum physics, and parallels the results obtained for 
simpler sources.
One new feature however emerges, in that this probability shows a modulation
with the momenta of the shells and the radius at which the shells collide, as a manifestation
of quantum mechanical interference.
Finally, we also consider the case of one light shell collapsing into a larger central mass.
\par
\null
\par
%\textit{PACS - ...}
\end{abstract}
\section{Introduction}
The general relativistic study of the gravitational collapse leading to the formation
of black holes dates back to the seminal papers of Oppenheimer and co-workers~\cite{OS},
nonetheless it remains one of the most challenging issues of contemporary theoretical physics.
The literature has grown immensely~\cite{joshi}, but many technical and conceptual difficulties
remain unsolved, in particular when one wants to develop a quantum description of this process.
\par
What is unanimously accepted is that the gravitational force becomes dominat whenever
a large enough amount of matter is localized within a sufficiently small volume.
Thorne captured the essence of black hole formation from two colliding objects in what
is known as the {\em hoop conjecture\/}~\cite{Thorne:1972ji}, which roughly states that
a black hole will form when the impact parameter $b$ is shorter than the
Schwarzschild radius $\rh$ of the system, that is for
\be
b
\lesssim
2\,\lp\,\frac{E}{\mpl}
\equiv
\rh
\ ,
\label{hoop}
\ee
where $E$ is total energy in the centre-of-mass frame.
Note that we use units with $c=1$, the Newton constant $\gn=\lp/\mpl$, where $\lp$ and $\mpl$ are
the Planck length and mass, respectively, and $\hbar=\lp\,\mpl$.
The main advantage of these units is to make it apparent that the Newton constant
converts mass into length (or the other way around) and provides a natural link
between energy and position (as we shall make more explicit in Section~\ref{idea}).
Initially formulated for black holes of astrophysical size~\cite{payne}, for which the concept
of a classical background metric and related horizon structure should be reasonably safe,
the hoop conjecture has now been analysed theoretically for a variety of situations.
\par
One of the most important questions which arise is whether the above conclusion
works the same way when the colliding masses (to be more specific, the total energy of the system)
approach down to the Planck scale.
Answering this question is extremely difficult because quantum effects may hardly be neglected
(see, e.g.~Ref.~\cite{acmo}) and it cannot be excluded that the purely general relativistic picture
of black holes must be replaced in order to include the possible existence of new Planck size objects,
generically referred to as ``quantum black holes'' (see, e.g.~Refs.~\cite{hsu,qbh}).
The challenge, when dealing with quantum black holes is to describe a system containing
quantum mechanical objects (such as the elementary particles of the Standard Model) and,
at the same time, identify the presence of horizons.
\par
It was recently proposed in Ref.~\cite{Casadio:2013tma} to define a wave-function for the
horizon (HWF) which can be associated with any localised quantum mechanical particle described
by a wave-function in position space.
This Horizon Quantum Mechanics (HQM) precisely serves the purpose to compute the probability
of finding the horizon of a certain radius centred around the source.  
Following this prescription one can directly associate to each quantum mechanical particle
a probability that it is a black hole.
In most cases, such a probability is a rather steep function of the energy which decreases to zero
quite rapidly below the Planck scale.
One consequently finds that there effectively exists a minimum black hole mass, albeit not in the
form of a sharp threshold, which entails expectations both from the classical hoop conjecture and the
Heisenberg uncertainty principle of quantum physics.
Further developments of this proposal can be found in
Refs.~\cite{Casadio:2013aua,Casadio:2013uga,Casadio:2015rwa,Casadio:2015qaq,Casadio:2016fev,
Casadio:2015jha,Casadio:2017nfg,Giugno:2017xtl,Casadio:2018mry}.
\par
Thin spherically symmetric layers of matter, or shells, are a very common toy model to investigate the
classical dynamics of the gravitational collapse in general relativity (see,
e.g.~Refs.~\cite{Kijowski:2005qm,Cardoso:2016wcr} and references theorein).
In this work we will generalise the HQM to the case in which the matter source consists of 
spherically symmetric and concentric shells, in their centre-of-mass frame, and still neglecting the
time evolution. 
Our aim is in particular to analyse the collision of concentric shells and study the probability that
a horizon forms by deriving the HWF of the system at the moment the shells collide.
For this purpose, we shall describe the quantum state of each shell as a Gaussian wave-function
in position space and further take the ultra-relativistic limit of very large radial momentum
(compared to the shell proper mass).
A new effect will emerge, in the form of a modulation of the probability that a system of two
shells is a black hole, as a straightforward consequence of the shell wave-functions being
complex in momentum space.
\par
The paper is organised as follows:
in Section~\ref{idea}, we shall briefly review the HQM for a single spherically symmetric source and
generalise it to the case of $N$ concentric shells.
Since it is in general impossible to obtain analytical results, suitably approximate equations for the case
of two shells are obtained in Section~\ref{particular2}, where we will also analyse several different configurations,
including the case of a single shell collapsing into a much heavier central source;
conclusions and future perspectives are summarised in the final Section~\ref{S:conc}.
\section{Horizon Quantum Mechanics for spherical systems}
\label{idea}
In this section we first review the basics about the idea of an auxiliary HWF to describe
the gravitational radius of a quantum state for a single particle, and then generalise it to the case of
$N$ concentric shells.
\subsection{Single particle case}
\label{SPC}
As we noted in the Introduction, Newton's constant naturally relates mass and length
and can therefore be used to define a HWF given the quantum mechanical
wave-function of a particle in position space.
This idea was first put forward in Ref.~\cite{Casadio:2013tma} and more details about its 
mathematical formulation can be found in Ref.~\cite{Casadio:2016fev}.
\par
In a spherically symmetric space-time, the line element can always be written as
\be
\d s^2
=
g_{ij}\,\d x^i\,\d x^j
+
r^2(x^i)\left(\d\theta^2+\sin^2\theta\,\d\phi^2\right)
\ ,
\label{metric}
\ee
with $x^i=(x^1,x^2)$ coordinates on surfaces where the angles $\theta$ and $\phi$
are constant.
The location of a trapping horizon is then determined by 
\be
0
=
g^{ij}\,\nabla_i r\,\nabla_j r
=
1-\frac{2\,M}{r}
\ ,
\label{th}
\ee
where $\nabla_i r$ is the covector perpendicular to surfaces of constant area
$\mathcal{A}=4\,\pi\,r^2$, and $M=\lp\,m/\mpl$ is the active gravitational (or Misner-Sharp)
mass, representing the total energy enclosed within a sphere of area $\mathcal{A}$.
If we set $x^1=t$ and $x^2=r$, the function $m$ is explicitly given by 
\be
m(t,r)=\frac{4\,\pi}{3}\int_0^r \rho(t, \bar r)\,\bar r^2\,\d \bar r
\ ,
\label{M}
\ee
where $\rho$ is the energy density of the matter source in the Einstein equations.
If we further assume the system is static, Eq.~\eqref{th} then simply identifies the horizon 
as the sphere of radial coordinate
\be
\rh
=
2\,M
\ ,
\label{Rh}
\ee
which becomes the usual expression of the Schwarzschild radius when we take the limit
$r\to\infty$ in which $m$ becomes the total ADM mass.
\par
The purpose of the HQM is to lift the condition~\eqref{th} 
(or, equivalently, the classical Eq.~\eqref{Rh}) to a quantum constraint that must be satisfied
by the physical states.
Let us then consider a wave-function $\psi_{\rm S}=\psi_{\rm S}(r)$ representing a spherically
symmetric object which is both localised in space and at rest in the chosen reference frame,
that is a ``particle'' of rest mass $m$.
This wave-function can be decomposed into energy eigenstates,
\be
\ket{\psi_{\rm S}}
=
\sum_E\,C(E)\,\ket{\psi_E} \label{spectral}
\ ,
\ee
where the sum represents the spectral decomposition in Hamiltonian eigenmodes,
\be
\hat H\,\ket{\psi_E}=E\,\ket{\psi_E}
\ ,
\ee
and $H$ can be specified depending on the model we wish to consider.
We then invert Eq.~\eqref{Rh} to obtain $E=\mpl\,M/\lp$ as a function of $\rh$, and
define the HWF as
\be
\psi_{\rm H}(\rh)
\propto
C\!\left(\mpl\,{\rh}/{2\,\lp}\right) 
\ ,
\ee
whose normalisation is finally fixed in the scalar product
\be
\pro{\psi_{\rm H}}{\phi_{\rm H}}
=
4\,\pi
\int_0^\infty
\psi_{\rm H}^*(\rh)\,\phi_{\rm H}(\rh)\,\rh^2\,\d \rh
\ .
\ee
We interpret the normalised wave-function $\psi_{\rm H}$ simply as yielding the probability
that $r=\rh$ is the gravitational radius associated with the particle in the given quantum state
$\psi_{\rm S}$.
The localisation of the horizon will consequently be governed by the uncertainty relation, like the position
of the particle
itself~\cite{Casadio:2013aua,Casadio:2013uga,Casadio:2015rwa,Casadio:2015qaq,Casadio:2015jha,
Casadio:2016fev,Casadio:2017nfg,Giugno:2017xtl,Casadio:2018mry}.
\par
Having defined the $\psi_{\rm H}$ associated with a given $\psi_{\rm S}$,
we can now compute the probability that the particle is a black hole as
\be
P_{\rm BH}
=
\int_0^\infty 
{\mathcal P}_<(r<\rh)\,\d \rh
\ ,
\label{PBH}
\ee
where
\be
{\mathcal P}_<(r<\rh)
=
P_{\rm S}(r<\rh)\,{\mathcal P}_{\rm H}(\rh)
\label{PrlessH}
\ee
is the probability density that the particle lies inside its own gravitational radius $r=\rh$.
The latter is in turn determined by the product of the probability that the particle is found inside
a sphere of radius $r=\rh$,
\be
P_{\rm S}(r<\rh)
=
4\,\pi\,\int_0^{\rh}
|\psi_{\rm S}(r)|^2\,r^2\,\d r
\ ,
\label{PsRh}
\ee
and the probability density for the horizon to be located on the sphere of radius $r=\rh$,
\be
{\mathcal P}_{\rm H}(\rh)
=
4\,\pi\,\rh^2\,|\psi_{\rm H}(\rh)|^2
\ .
\label{PhRh}
\ee
\subsection{Concentric Gaussian shells}
\label{general}
We now proceed to apply the previous formalism to the case of a system composed of $N$ shells
with common centre and different radii as well as shrinking (or expanding) velocities.
The size of each shell will be described by a Gaussian wave-function in position space and,
since the HQM for time-dependent systems has not yet been fully developed
(see Ref.~\cite{Casadio:2014twa}), we shall here just consider ``snap-shots'' of the system at
given instants of time, like in Ref.~\cite{Casadio:2013uga}.
\par
Let us denote with $m_a$, $R_a$ and $v_a$ the masses, areal radii and velocities of 
expansion or contraction of the shells, respectively, where $a=1,\ldots,N\ge 2$.
Radii and velocities will in general vary in time, but since we are going to compute the
HWF for the system at a given instant of time, we can treat those as constants.
In this respect, we are considering a simplified version of the system of many nested shells
studied in Refs.~\cite{Nshells}.
For further simplicity, and differently from Refs.~\cite{Nshells}, the background metric is
assumed to be flat, although corrections could be derived for a Gaussian distribution of
classical energy along the lines of Refs.~\cite{nicolini,bec}.
\par
According to the usual quantum mechanical prescription, the wave-function of a system of $N$
shells is given by the product~\footnote{We do not (anti)symmetrise the product
of the wave-functions since the shells may have different masses and might therefore be
distinguishable.} 
\be
\psi_{\rm S}(r_1,\ldots,r_N)
=
\prod_{a=1}^N\,\psi_{\rm S}(r_a;\ell_a,R_a,P_a)
\ .
\label{psi_a_prod}
\ee
We assume the individual wave-functions are spherical waves with a Gaussian profile,
\be
\psi_{\rm S}(r_a;\ell_a,R_a,P_a)
=
N_a\,e^{-\strut\displaystyle\frac{\left(r_a-R_a\right)^2}{2\,\ell_a^2}}
\,
\frac{e^{\strut\displaystyle i\,\frac{P_a\,r_a}{\hbar}}}{r_a}
\ ,
\label{psi_a}
\ee
with $\ell_a$ the (Lorentz contracted) width of the shell, that is
\be
\ell_a
=
\sqrt{1-v_a^2}\,\bar\ell_a
=
\frac{m_a\,\bar\ell_a}{\sqrt{P_a^2+m_a^2}}
\ ,
\label{ell_a}
\ee
where we shall often assume the width of the shell at rest is given by the Compton
relation
\be
\bar\ell_a
\simeq
\frac{\hbar}{m_a}
=
\frac{\lp\,\mpl}{m_a}
\ .
\label{Comp}
\ee
Finally, the normalisation factor
\be
N_a
=
\left\{
2\,\pi^{3/2}\,\ell_a
\left[1+\text{Erf}\left(\frac{R_a}{\ell_a}\right)\right]
\right\}^{-1/2}\label{N_a}
\ee
ensures that
\be
4\,\pi
\int_0^{\infty}
\left|\psi_{\rm S}(r_a;\ell_a,R_a,P_a)\right|^2
r_a^2\,d r_a
=
1
\ ,
\ee
for all $a=1,\ldots,N$.
\par
For each of the above wave-functions we have 
\be
\expec{r_a}
=
R_a
+
\frac{\ell_a\,e^{-\frac{R_a^2}{\ell_a^2}}}
{\sqrt{\pi}\left[1+\text{Erf}\!\left(\frac{R_a}{\ell_a}\right)\right]}
\simeq
R_a
\ ,
\ee
where the approximation holds for $R_a\gg \ell_a$.
Likewise, the expectation value for the radial momentum, which will be used to calculate the energy of the shells,
is
\be
\expec{p_a^2}
\simeq
P_a^2
+
\frac{\mpl^2\,\lp^2}{2\,\ell_a^2}
\ .
\label{expec_p}
\ee
\par
Linearity of the spectral decomposition allows us to expand the wave-function of each shell
in energy eigenstates and then add the results, which we can formally write as
\be
\tilde\psi_{\rm S}(E)
=
\sum_{E=\sum E_a}
\left[
\prod_{a=1}^N\,\psi_{\rm S}(E_a;\ell_a,R_a,P_a)
\right]
\ ,
\label{psiSE}
\ee
where the superposition depends on the choice of spectral modes.
Since we are considering a spherically symmetric system, we have
a natural choice given by the eigenmodes of the spatial Laplacian,
\be
-\frac{\hbar^2 }{r^2}\,\frac{\partial}{\partial r}
\left(r^2\,\frac{\partial}{\partial r}\right)
\Psi_p
=
p^2\,\Psi_p
\ ,
\ee
that is, the spherical Bessel function of degree zero 
\be
\Psi_p
=
j_0(r\,p/\hbar)
=
\frac{\hbar \, \sin(r\,p/\hbar)}{r\,p}
\ .
\label{j0}
\ee
Since $j_0(-z)=j_0(z)$, we can always assume $p>0$, and the momentum eigenmodes
satisfy the condition 
\be
4\,\pi
\int_0^\infty
j_0(p\,r/\hbar)\,j_0(q\,r/\hbar)\,r^2\,\d r
=
\frac{2\,\pi^2\, \hbar^3 }{p^2}\,
\delta(p-q)
\ ,
\label{j00pqp}
\ee
which holds when both $p$ and $q>0$.
\par
We then have
\be
\tilde\psi_{\rm S}(p_a;\ell_a,R_a,P_a)
=
\sqrt{\frac{2}{{\pi\,\hbar^3}}}
\int_0^\infty
\psi_{\rm S}(r;\ell_a,R_a,P_a)\,j_0(p_a\, r/\hbar)\,
r^2\,\d r
\ ,
\label{fourier}
\ee
and we obtain
\be
\tilde\psi_{\rm S}(p_a;\ell_a,R_a,P_a)
&\!\!\!=\!\!\!&
\frac{i\,\sqrt{2}\,N_a}{\sqrt{\hbar}}
\int_0^\infty
e^{-\frac{\left(r-R_a\right)^2}{2\,\ell_a^2}}\,
{\left[e^{-i\,\frac{(p_a-P_a)\,r}{\hbar}}-e^{i\,\frac{(p_a+P_a)\,r}{\hbar}}\right]}
\,\d r
\nonumber
\\
&\!\!\!=\!\!\!&
\frac{i\,\ell_a\, N_a}{2\, \sqrt{\hbar}}
\left\{
\frac{e^{-i\,\frac{R_a\left(p_a-P_a\right)}{\hbar}}}{p_a}\,
e^{-\frac{\ell_a^2 \left(p_a-P_a\right)^2}{2\,\hbar^2}}
\left[1 + 
\Erf\!\left(\frac{R_a}{\sqrt{2}\, \ell_a}
+i\,\frac{\ell_a\left(P_a-p_a\right)}{\sqrt{2}\, \hbar}
\right) \right]
\right.
\nonumber
\\
&&
\left.
-
\frac{e^{+i\,\frac{R_a\left(p_a+P_a\right)}{\hbar}}}{p_a}\,
e^{-\frac{\ell_a^2\left(p_a+P_a\right)^2}{2\,\hbar^2}}
\left[1 
+\Erf\! \left(\frac{R_a}{\sqrt{2}\, \ell_a}
+i\,\frac{\ell_a \left(P_a+p_a\right)}{\sqrt{2}\, \hbar}
\right) \right]
\right\}
\ .
\qquad
\label{fourier_gaussian}
\ee
We then notice that, since $p_a>0$, the main contribution comes from the Gaussian
centred around $\bar P_a\equiv |P_a|$ and we can approximate the above expressions simply as
\be
\tilde\psi_{\rm S}(p_a;\ell_a,R_a,\pm \bar P_a)
&\!\!\simeq\!\!&
\pm \bar N_a\,
\frac{e^{\strut\displaystyle\mp i\,\frac{R_a\,(p_a-\bar P_a)}{\lp\,\mpl}}}{p_a}\,
e^{-\strut\displaystyle\frac{\ell_a^2\left(p_a-\bar P_a\right)^2}{2\,\lp^2\,\mpl^2}}
\ ,
\label{fourier_gaussian_approx}
\ee
where
\be
\bar N_a 
\simeq
\frac{i\, \ell_a\, N_a}{2\, \sqrt{\lp\,\mpl}}\left[1+\Erf \left(\frac{R_a}{\sqrt{2}\, \ell_a}\right)
\right] ,
\ee
with $N_a$ defined in Eq. \eqref{N_a}. 
\par
We can now use 
\be
\tilde\psi_{\rm S}(p_1,...,p_N)
=
\prod_{a=1}^N\,
\tilde\psi_{\rm S}(p_a;\ell_a,R_a,P_a)\ ,
\ee
and the dispersion relation
\be
E_a^2
=
p_a^2+m_a^2
\ ,
\label{dispersion_rel}
\ee
along with Eq.~\eqref{Rh}, in order to derive the HWF.
\section{Black hole probability for two shells}
\label{particular2}
In this section we consider several special cases in which the system is simply made of $N=2$ shells,
which could be of particular interest to investigate the horizon formation caused by the collision
of very thin relativistic layers of matter.
\par
The wave-function of each shell will be given by Eq.~\eqref{psi_a} with $a=1,2$, so that the total 
wave-function of the system of two shells is the direct product
\be
\psi_{\rm S}(r_1, r_2)
=
\prod_{a=1}^2\,\psi_{\rm S}(r_a;\ell_a,R_a,P_a)
\ .
\ee
In order to compute the spectral decomposition, we go through momentum space 
and employ the approximate expression~\eqref{fourier_gaussian_approx} for each shell.
The two-shell state can then be written as
\be
\ket{\psi_{\rm S}^{(1,2)}}
=
\prod_{a=1}^2
\left[
4\, \pi
\int\limits_{0}^{\infty}
p_a^2\,\d p_a\,\tilde\psi_{\rm S}(p_a;\ell_a,R_a,P_a)\,\ket{p_a}
\right]
\ .
\label{PsiPp}
\ee
The relevant coefficients in the spectral decomposition~\eqref{spectral} are given by the sum~\eqref{psiSE}
of all the components of the product wave-function with the same total energy $E$, that is
\be
C(E)
&\!\!=\!\!&
\bra{E}
\int_0^\infty
\d E'\,\ket{E'}\bra{E'}
\prod_{a=1}^2
\left[
4\, \pi
\int\limits_{0}^{\infty}
p_a^2\,\d p_a\,\tilde\psi_{\rm S}(p_a;\ell_a,R_a,P_a)\,\ket{p_a}
\right]
\nonumber
\\
&\!\!=\!\!&
16\, \pi^2
\int\limits_{0}^{\infty}
\int\limits_{0}^{\infty}
\tilde\psi_{\rm S}(p_1)\,\tilde\psi_{\rm S}(p_2)\,
\delta(E-E_1(p_1)-E_2(p_2))\,p_1^2 \, p_2^2\,
\d p_1\d p_2
\ ,
\label{C(E)}
\ee
with $p_a$ and $E_a$ related by the relativistic dispersion relation \eqref{dispersion_rel}. 
\par
Assuming that the rest masses $m_a$ of the shells are much smaller than the Planck scale, black holes
are expected to form with a significant probability only when the momenta of the two shells
are of the order of the Planck mass, that is $|P_a|\gg m_a$. 
Eq.~\eqref{expec_p} then yields 
\be
\expec{p_a^2}
\simeq
2\, P_a^2
\ .
\ee
and, if we employ the Compton relation~\eqref{Comp}, the Lorentz contracted width~\eqref{ell_a} of the
shells becomes
\be
\ell_a
\simeq
\frac{\lp\,\mpl}{\sqrt{\expec{p_a^2}+m_a^2}}
\sim
\frac{\lp\,\mpl}{\sqrt{2}\,|P_a|}
\ .
\ee 
\par
We next consider different combinations of $R_a$ and $P_a$ and compute the corresponding
probabilities~\eqref{PBH} that they are black holes.
\subsection{Shells collapsing with equal speeds at same radius}
\label{S:2equal}
We first consider two shells of equal mass travelling together with equal radial velocities,
$v_1=v_2\equiv -v$, at the moment their radii $R_1=R_2\equiv R>0$.
Given their equal masses, the two shells also have the same momenta $P_1=P_2 \equiv -P$ and equal Lorentz
contracted widths.
The case of two shells overlapping at zero mean radius ($R=0$) will also be discussed at the end of this
subsection.  
\par
The wave-functions for the two shells are described in position space by Eq.~\eqref{psi_a}, while the
wave-function of the system is the product of the two, as shown in Eq.~\eqref{psi_a_prod}.
In momentum space the wave-functions of the two shells are given by the expressions in
Eq.~\eqref{fourier_gaussian}, which simplify to the corresponding approximate expressions in
Eq.~\eqref{fourier_gaussian_approx}.
In particular, since we are considering collapsing shells, the momenta are negative and one needs
to use the lower signs in Eq.~\eqref{fourier_gaussian_approx}.
\subsubsection{Finite mean radius}
As detailed previously, we use the momentum space wave-functions to compute the unnormalized
HWF by replacing the expression for the Schwarzschild radius from Eq.~\eqref{hoop}
into Eq.~\eqref{C(E)} and obtain
\be
\psi_{\rm H}(\rh)
&\!\!\propto\!\!&
\frac{\left[\Erf\!\left(\frac{R \,P}{\sqrt{2}\,\lp\mpl}\right)+1\right]^2}
{\Erf\!\left(\frac{R \,P}{\lp\mpl}\right)+1}\,
\exp\!\left(-\frac{\mpl^2\,\rh^2}{16\, \lp^2\,P^2}
+\frac{\mpl\,\rh}{2\, \lp\,P}
-\frac{i\,R \,\rh}{2\,\lp^2}
+2 \,i\, \frac{R \,P}{\lp\,\mpl}\right)
\nonumber
\\
&&
\times
\left[
\left(\frac{\rh^2}{\lp^2}-8 \,\frac{P^2}{\mpl^2}\right)
\Erf\!\left(\frac{\mpl\,\rh}{4\, \lp\,P}\right)
+
\frac{4\,e^{-\frac{\mpl^2\,\rh^2}{16\, \lp^2\,P^2}}}{\sqrt{\pi}}\,
\frac{\rh \,P}{\lp\,\mpl}
\right]
\ .
\label{HWF_same_P0}
\ee
\begin{figure*}[]
\centering
\includegraphics[width=12cm]{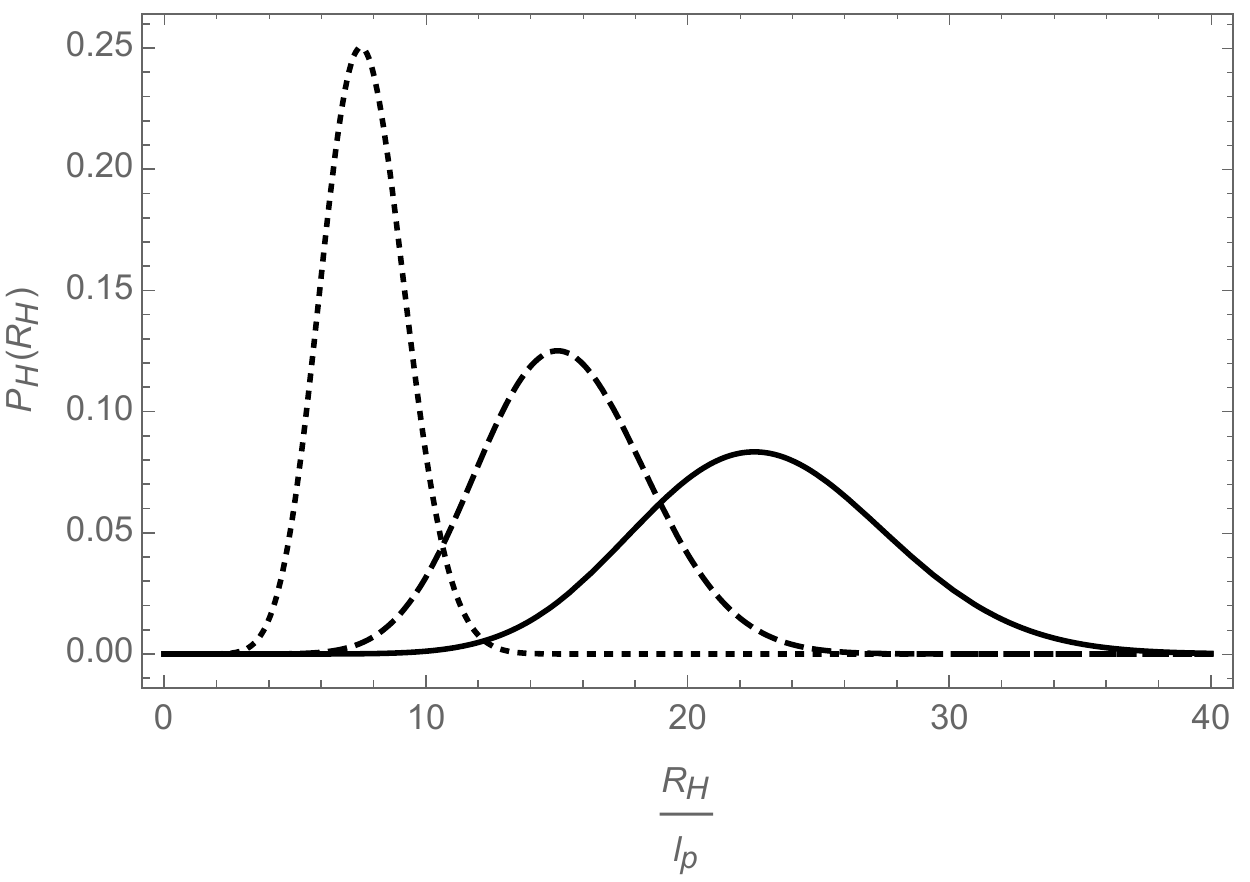}
\caption{Probability density for the horizon to be located on the sphere of radius $r=\rh$,
for a total energy $E=2\,\sqrt{2}\,\mpl\simeq 2.8\,\mpl$ (dotted line), $E=4\,\sqrt{2}\,\mpl\simeq 5.7\,\mpl$
(dashed line) and $E=6\,\sqrt{2}\,\mpl\simeq 8.5\,\mpl$ (continuous line). 
Curves peak at values larger than $\rh\sim2\,E$.}
\label{PsiPsi_same_P0}
\end{figure*}
\par
The probability density ${\mathcal P_{\rm H}}$ in Eq.~\eqref{PhRh} for the horizon to be located
on the sphere of radius $r=\rh$ is shown in Fig.~\ref{PsiPsi_same_P0}.
Unlike the cases previously
considered~\cite{Casadio:2013aua, Casadio:2013uga, Casadio:2015rwa, Casadio:2015qaq, Casadio:2017nfg}, 
this probability density becomes maximum at values of $\rh$ slightly larger than twice the total energy of the system. 
While the hoop conjecture~\eqref{hoop} suggests that the peak should be located around
$\rh\simeq 2\,\lp\,(2\,|\sqrt{\expec{p_a^2}}|/\mpl)=4\sqrt{2}\,\lp\,|P|/\mpl$, the probability
density peaks at values of $\rh$ corresponding to larger values of the total energy $E$, as it can be seen in
Fig.~\ref{PsiPsi_same_P0}.
For instance, when considering the case $P=\mpl$, the total energy should be equal to $2\sqrt{2}\,\mpl$
and the horizon radius of this system should be $\rh\simeq 4\sqrt{2}\,\lp$.
This case is represented by the dotted line in Fig.~\ref{PsiPsi_same_P0}, and we can see that
${\mathcal P_{\rm H}}$ is maximum around $\rh\simeq 8\,\lp$.
This is also true for the other two cases plotted in the same figure.
The values of the total energies are mentioned in the caption and one can easily verify that
in each instance the location of the peak is at a larger value than the one expected from Eq.~\eqref{hoop}. 
\begin{figure*}[]
\centering
\includegraphics[width=11cm]{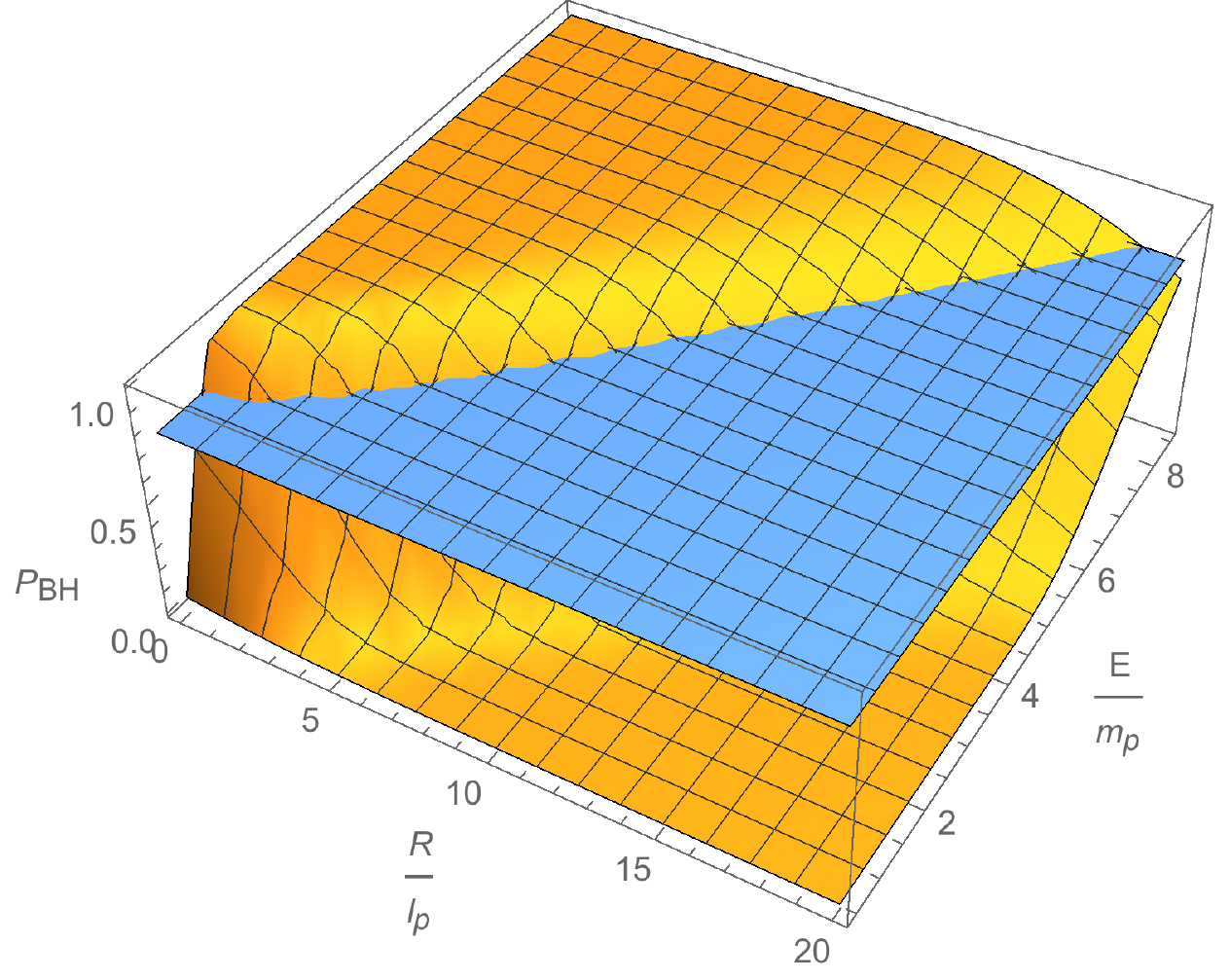}
\caption{Probability for two shells of equal radial momenta to be a black hole as a function of the radius $R$
and the total energy $E$ (in Planck units).
The blue plane delimits the region above which the probability $P_{\rm BH}>0.8$.}
\label{P_BH_same_P0}
\end{figure*}
\par
\begin{figure*}[t]
\centering
\includegraphics[width=7.5cm]{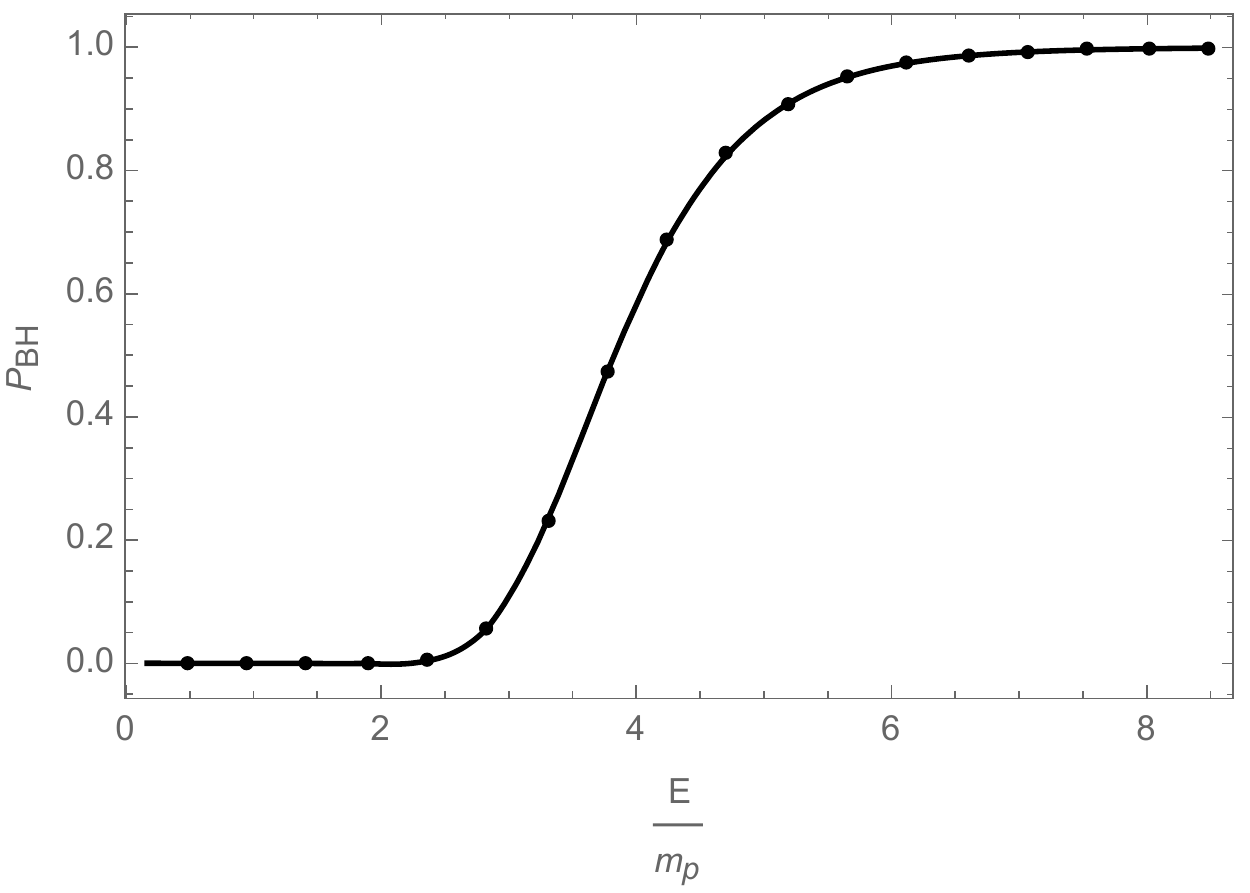}
\includegraphics[width=7.5cm]{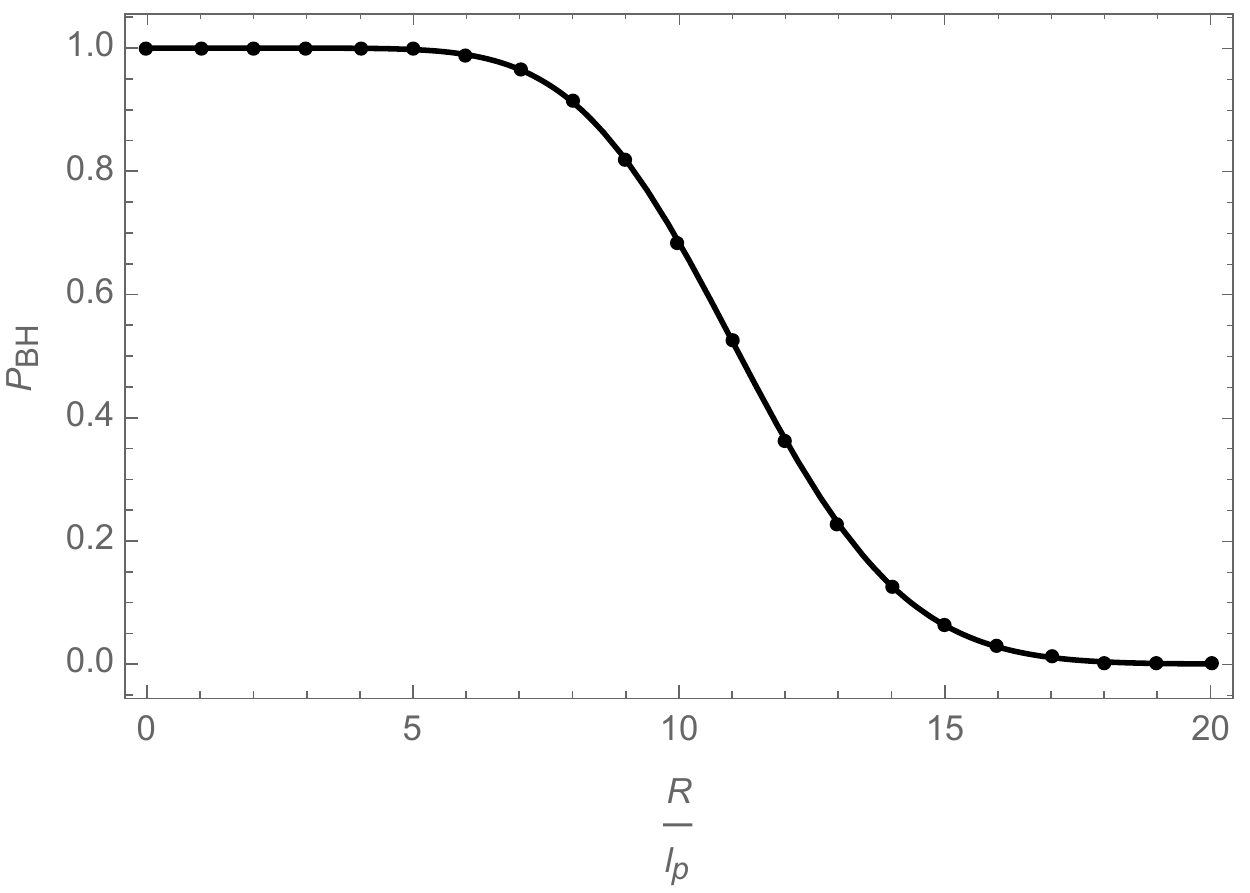}
\caption{Probability for the two-shell system to be a black hole as a function of the total
energy $E$ for $R=10\, \lp$ (left plot) and the same probability as a function of the mean
radius $R$ for $E= 3\,\sqrt{2}\, \mpl\simeq 4.2\,\mpl$ (right plot).
Dots represent the values computed numerically.}
\label{P_BH_slices}
\end{figure*}

After normalising the HWF, one can use it to calculate the probability
$P_{\rm BH}=P_{\rm BH}(R, E)=P_{\rm BH}(R, 2\sqrt{2}\,|P|)$ for the system of two shells to be
a black hole as a function of the radius of the shells $R$ and the value of the total energy of the system $E$,
by following the procedure described in the last part of Section~\ref{SPC}.
This probability is displayed by the three-dimensional plot in Fig.~\ref{P_BH_same_P0},
where the horizontal plane intersects the graph at $P_{\rm BH}=80\%$ for easy reference.
Two slices from this three-dimensional plot are displayed in Fig.~\ref{P_BH_slices}:
one graph represents the probability for the two shells to be a black hole as a function
of the total energy $E$ for a constant value of the mean radius $R$, while the other represents
the probability for the system to be a black hole as a function of the mean radius $R$
for a given value of the total energy $E$. 
\par
The plot on the left of Fig.~\ref{P_BH_slices}, obtained for a constant mean radius $R=10\, \lp$
(this should be understood in the sense that two shells collide at this particular radius)
shows that the probability $P_{\rm BH}$ is already rather large for values of the total energy
$E\simeq 4\, \mpl$, which is below the value of $E=5\, \mpl$ that one calculates from the classical
hoop conjecture for two shells overlapping at $R=10\, \lp$.
The classical hoop conjecture suggests the existence of a threshold effect for black hole formation
in the sense that these objects should only form when the impact parameter satisfies Eq.~\eqref{hoop}.
We see that the HQM instead predicts a smooth increase of the probability $P_{\rm BH}$ from zero
to one and, moreover, this probability is not zero for values of the total energy smaller than
the ones dictated by the hoop conjecture. 
\par
The same conclusion can be inferred from the plot on the right of Fig.~\ref{P_BH_slices},
which shows the dependence of the probability for the system of shells to be a black hole
as a function of the mean radius at which they collide for a total energy of the system
$E= 3\,\sqrt{2}\, \mpl$.
Again, the hoop conjecture would suggest that the probability for the system of shells to form
a black hole should drop to zero when they collide at mean radii values larger than
$6 \,\sqrt{2}\,\lp\simeq8.5\, \lp$.
Instead, the probability decreases slower and it is about $50\%$ at $R\simeq 11\, \lp$.
\subsubsection {Vanishing mean radius}
\label{R=0}
\begin{figure*}[]
\centering
\includegraphics[width=11cm]{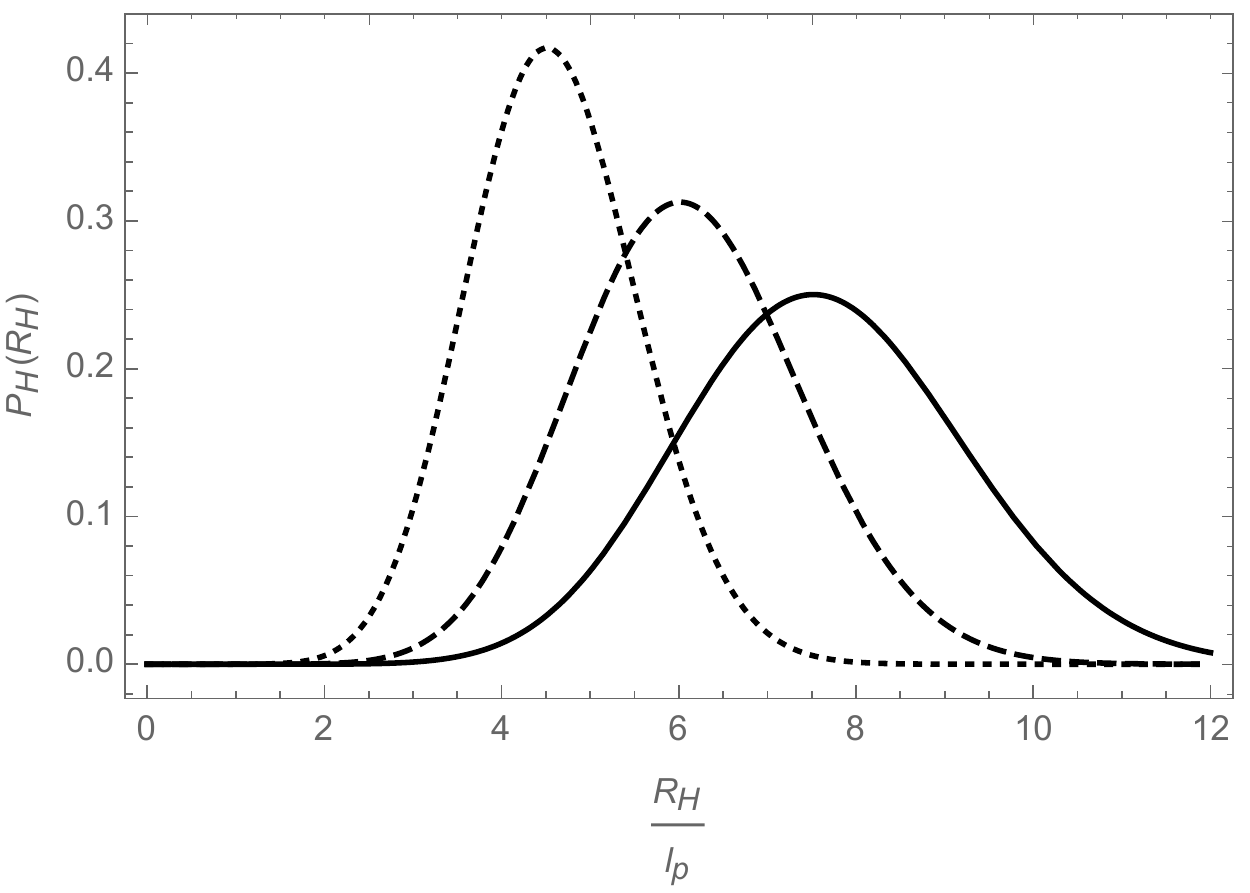}
\caption{Probability density $P_{\rm H}(\rh)=4\,\pi\,\rh^2\,|\psi_{\rm H}(\rh)|^2$ for the horizon to be
located on the sphere of radius $r=\rh$, for $E=1.2\,\sqrt{2}\,\mpl\simeq 1.7\,\mpl$ (dotted line),
$E=1.6\,\sqrt{2}\,\mpl\simeq 2.3\,\mpl$ (dashed line) and $E=2\,\sqrt{2}\,\mpl\simeq 2.8\,\mpl$
(continuous line) for two overlapping shells with vanishing mean radius. 
Curves peak at values larger than $\rh\sim2\,E$.}
\label{PsiPsi_same_P0_R=0}
\end{figure*}

A special case worth considering is when the two shells have reached zero mean radius.
The corresponding unnormalised HWF can be obtained by setting $R=0$
in Eq.~\eqref{HWF_same_P0}.
For two collapsing shells, this simplifies to
\be
\psi_{\rm H}(\rh)
\propto
{e^{-\frac{\mpl^2\,\rh^2}{16\, \lp^2\,P^2}+\frac{\mpl\,\rh}{2\, \lp\,P}}}
\left[
\left(8\, \frac{P^2}{\mpl^2}
-\frac{\rh^2}{\lp^2}\right) \Erf\left(\frac{\mpl\,\rh}{4\, \lp\,P}\right)
-e^{-\frac{\mpl^2\,\rh^2}{16\, \lp^2\,P^2}} \, \frac{4\,\rh \,P}{\sqrt{\pi}\,\lp\,\mpl}
\right]
\ .
\label{HWF_same_P0_R=0}
\ee
Similarly to the general case discussed earlier, the probability density for the horizon to be located
on the sphere of radius $r=\rh$ becomes maximum at values of $\rh$ larger than twice the energy
of the system, as can be seen in Fig.~\ref{PsiPsi_same_P0_R=0}.
\par
\begin{figure*}[]
\centering
\includegraphics[width=11cm]{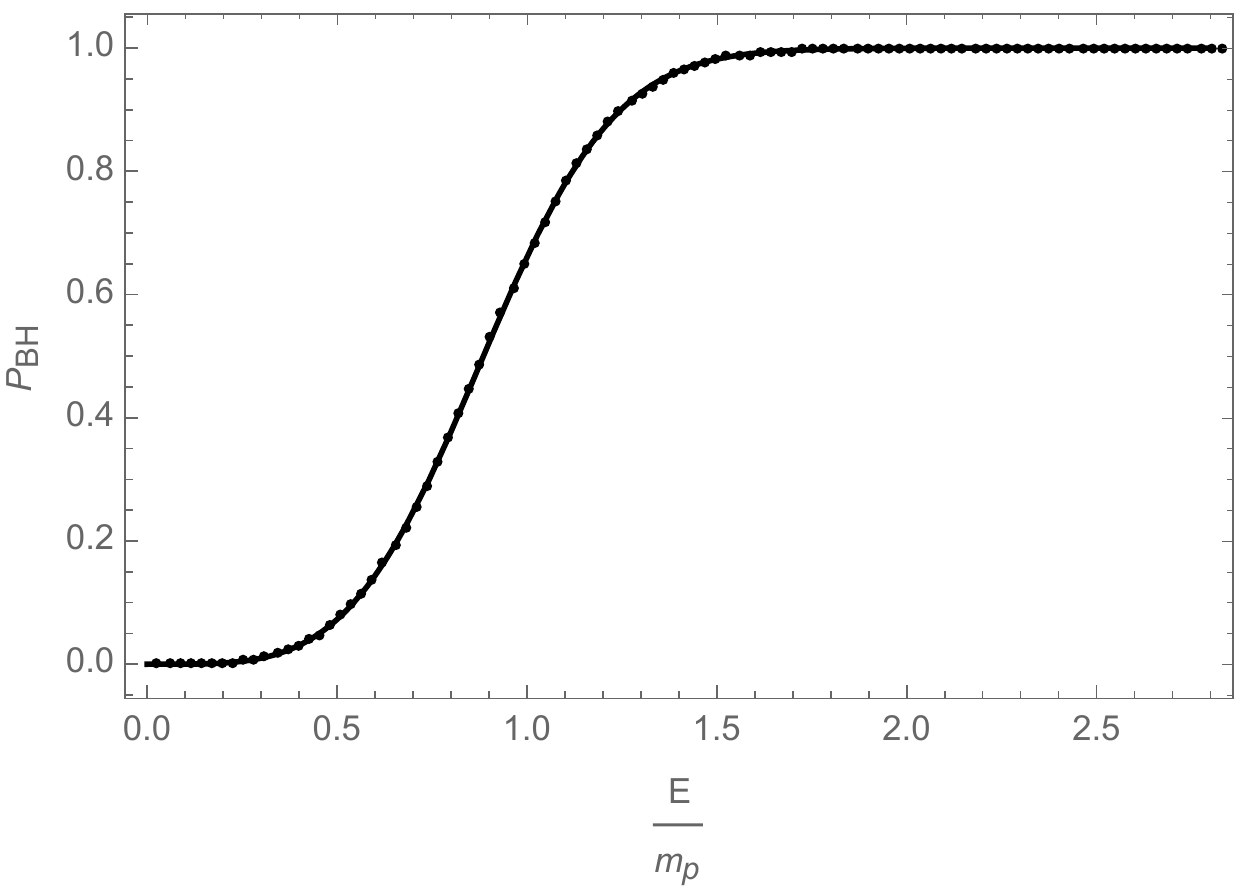}
\caption{Probability for the system of two shells that collide at $R=0$ to form a black hole as a function of
the total energy $E$ (in units of Planck mass).}
\label{P_BH_same_P0_R=0}
\end{figure*}

The plot in Fig.~\ref{P_BH_same_P0_R=0} shows the probability $P_{\rm BH}$ for the system of two
shells to be a black hole as a function of the the total energy of the system $E$.
This was, of course, obtained from the normalised HWF by following the procedure
described earlier.
The probability $P_{\rm BH}$ increases with the total energy $E$, reaching $50\%$ for
values of the total energy of about $\mpl$, which is also the threshold value suggested by the hoop
conjecture.
The plot shows that black holes can also form, with smaller probabilities, below this value. 
\subsection{Shells collapsing with different speeds at same radius}
\label{S:2different}
\begin{figure*}[]
\centering
\includegraphics[width=11cm]{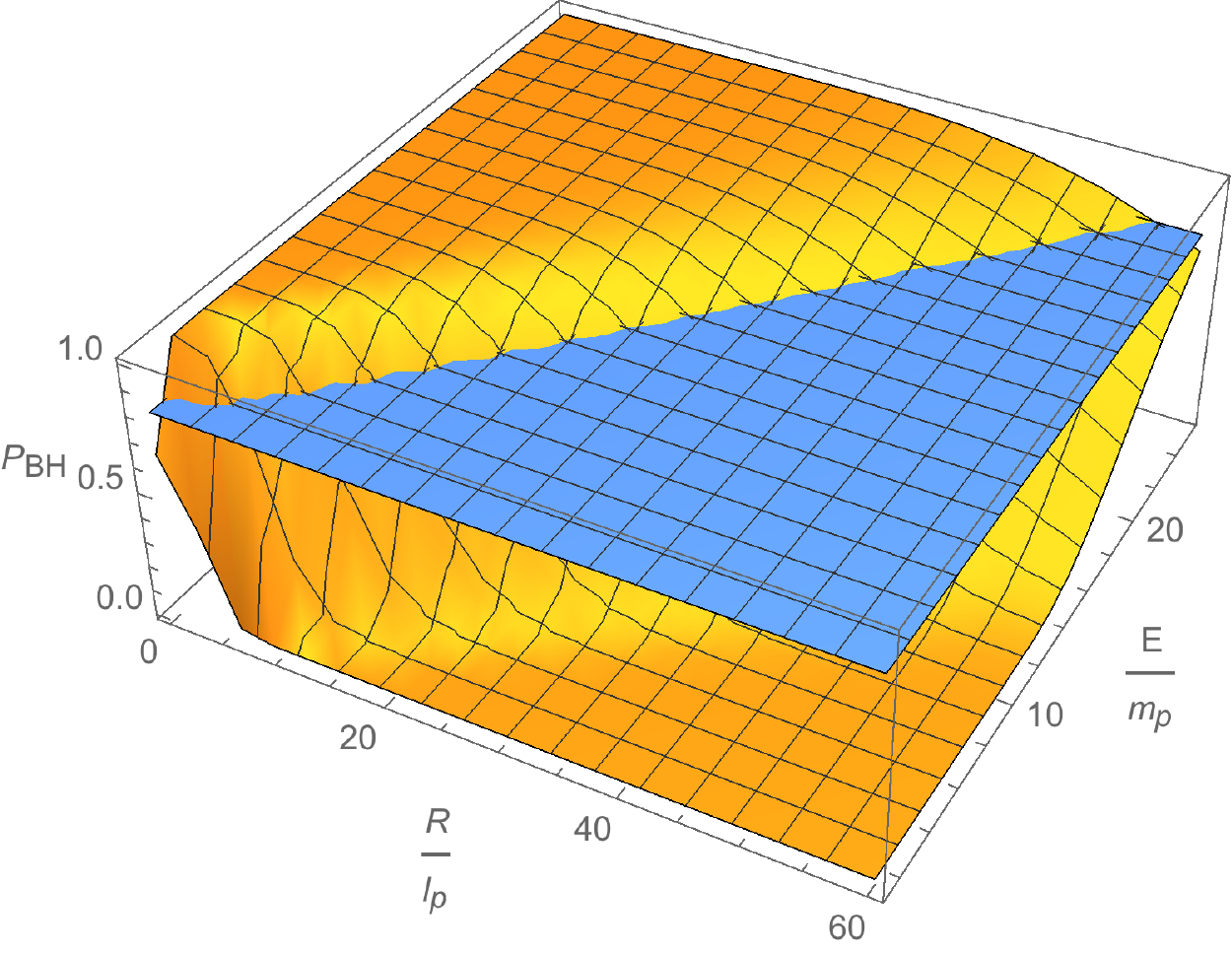}
\caption{Probability for two collapsing shells with momenta $P_1=-5\, P$ and $P_2=-P$ to form a black hole
as a function of the radius $R$ and the total energy $E$ (in Planck units).
The blue plane delimits the region where the probability $P_{\rm BH}>0.8$.}
\label{P_BH_different_P0}
\end{figure*}

A more general case is the one of two shells which have different momenta $P_1$ and $P_2$ that
collide at  $R_1=R_2\equiv R$.
The probability for the two shells to form a black hole as a result of the collision is calculated the same
way as earlier and the three dimensional plot of $P_{\rm BH}=P_{\rm BH}(R, E_1\,+\,E_2)$ for this case
is shown in Fig.~\ref{P_BH_different_P0}.
The values of the momenta for the two shells are $P_1=-5\, P$ and $P_2=-P$ (we remind our readers
that the negative signs mean that both shells are contracting). 
\par
\begin{figure*}[h!]
\centering
\includegraphics[width=7cm]{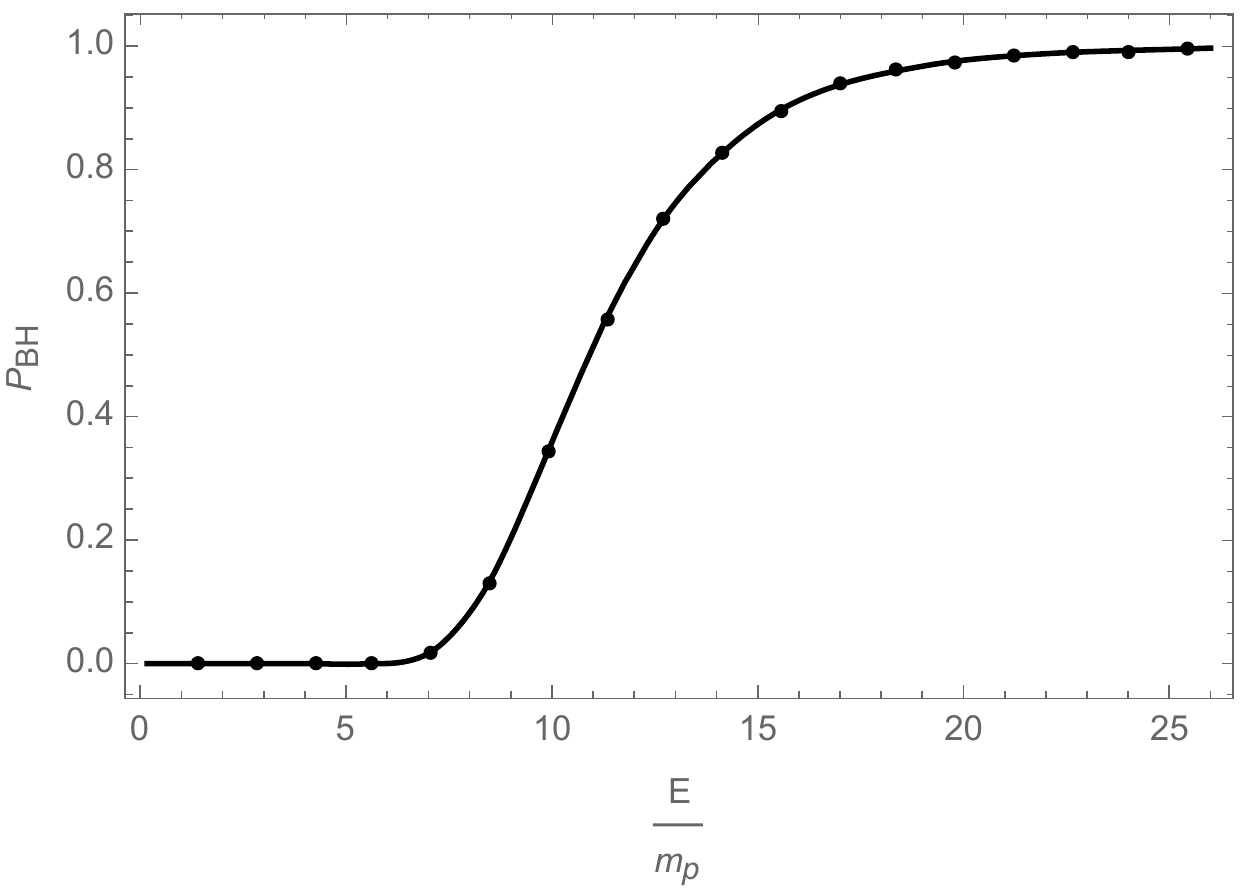}
\includegraphics[width=7cm]{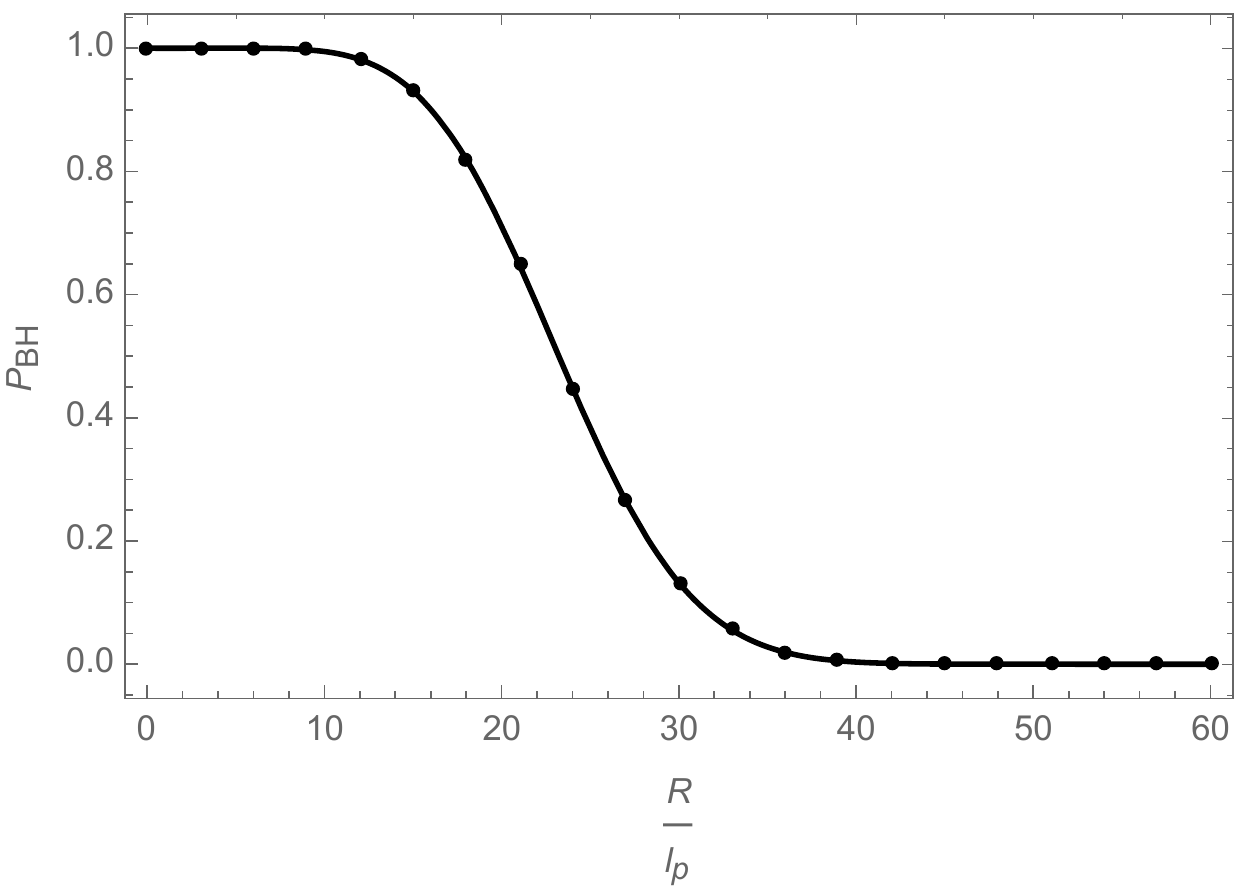}
\\
\includegraphics[width=7cm]{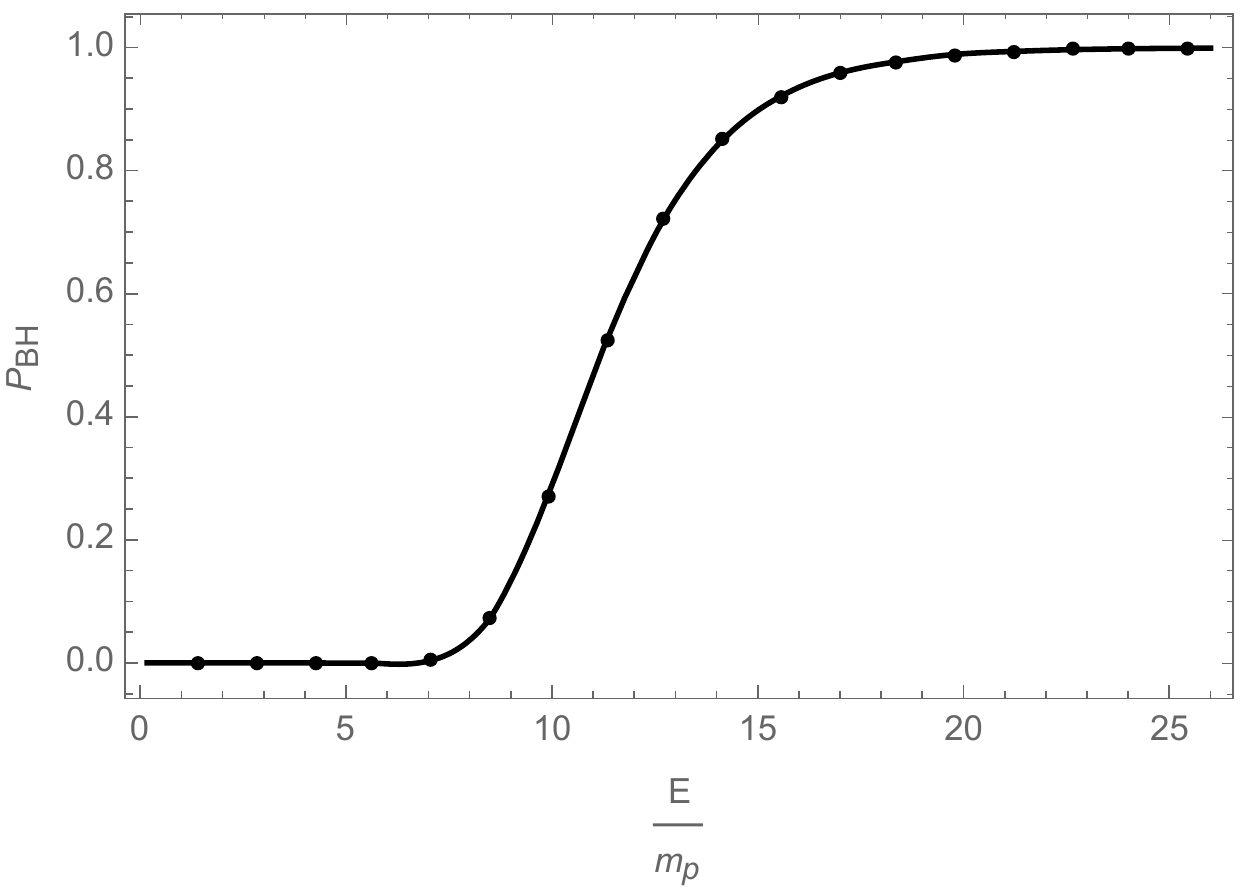}
\includegraphics[width=7cm]{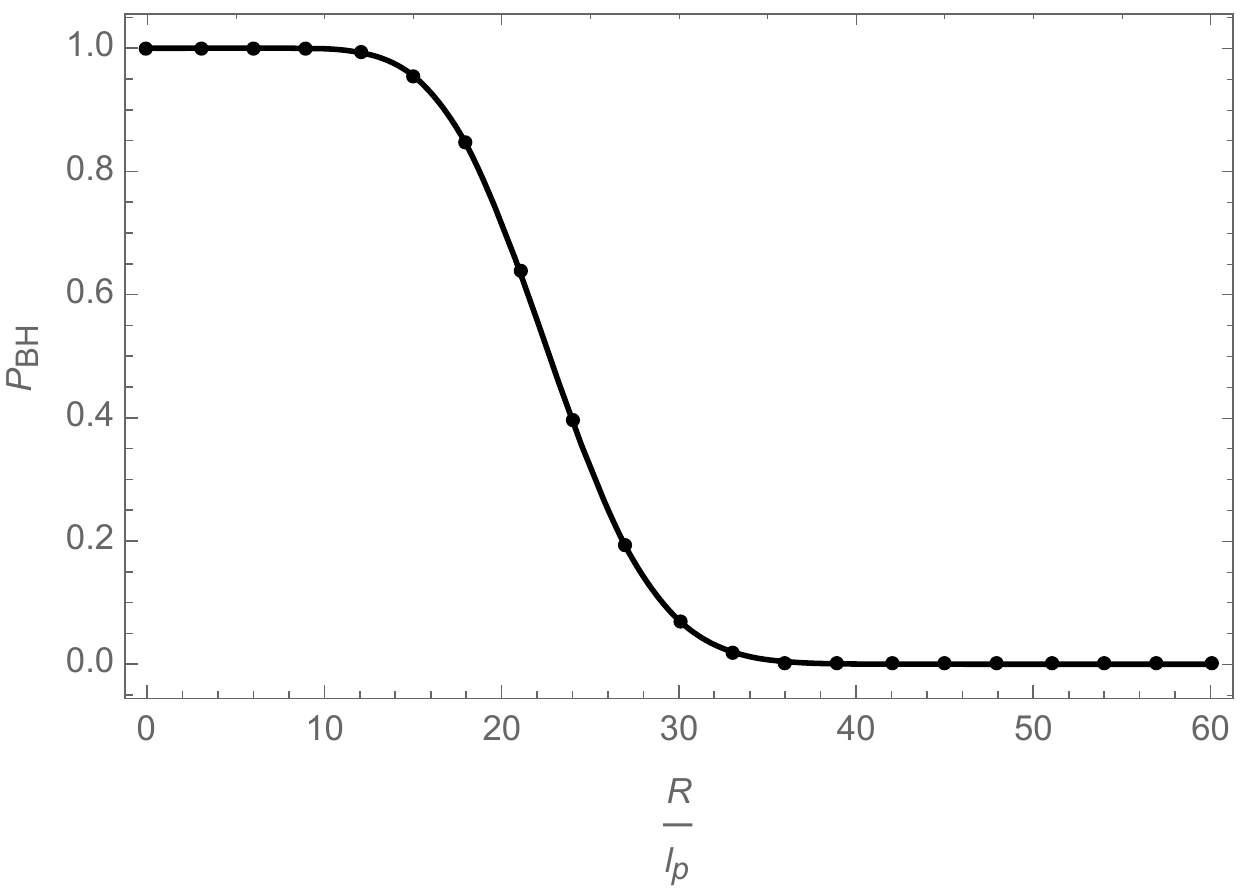}
\\
\includegraphics[width=7cm]{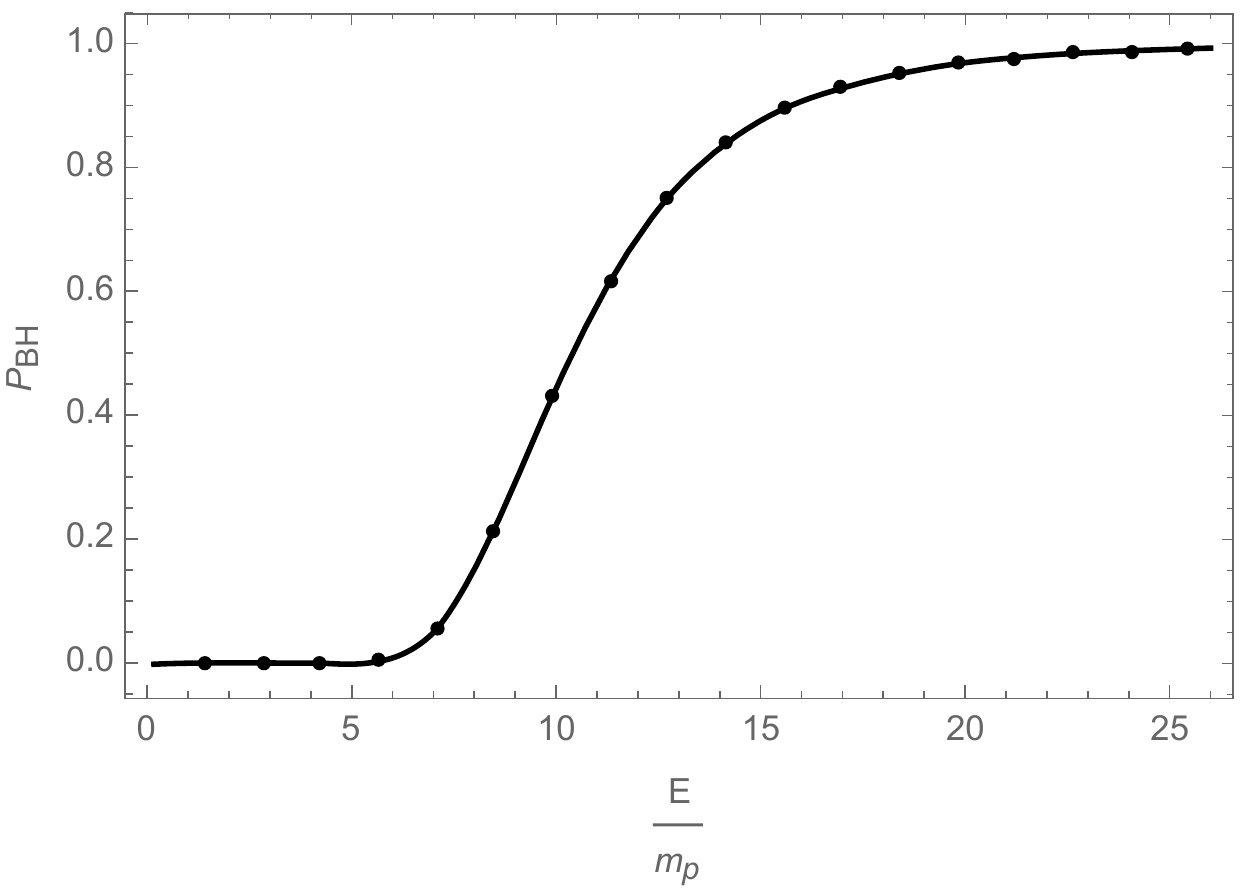}
\includegraphics[width=7cm]{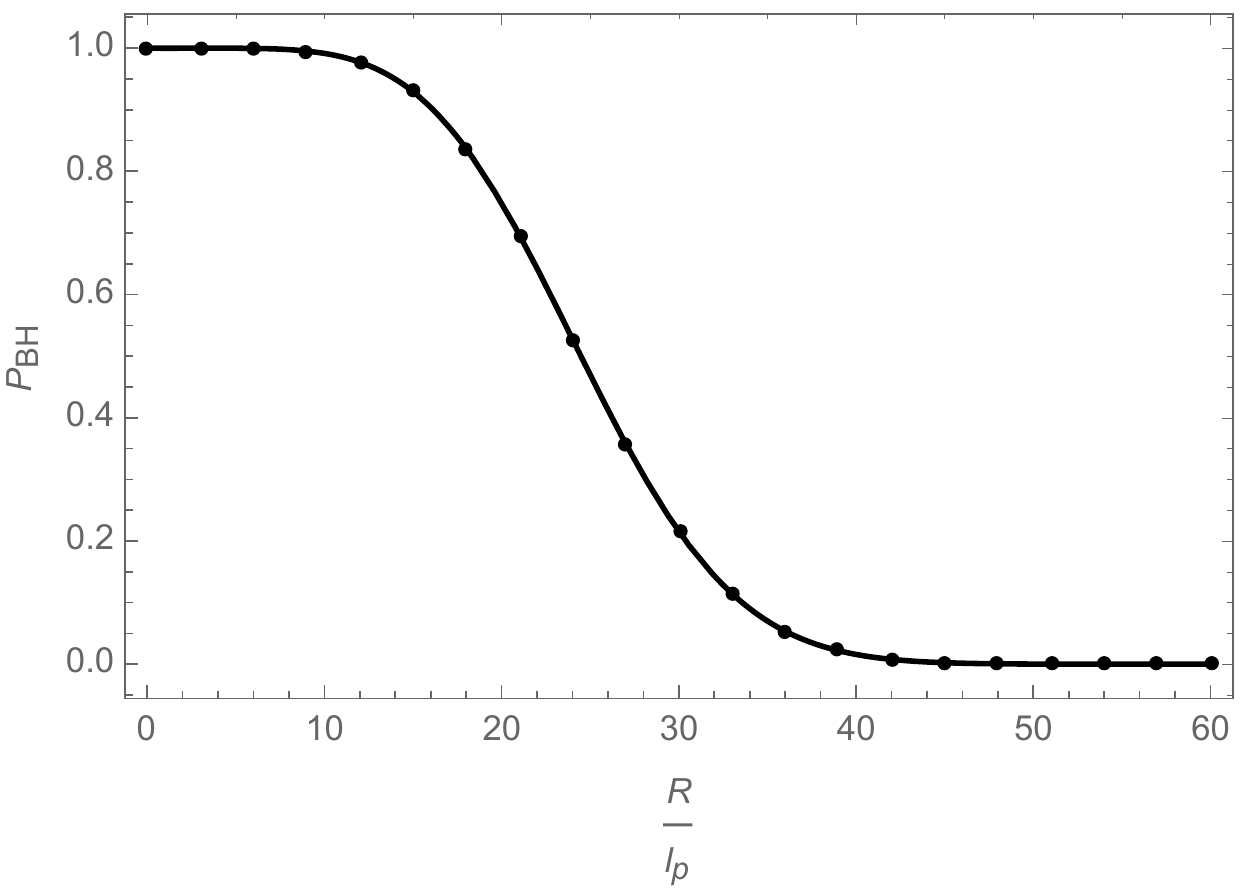}
\caption{\underline{Top left}:
Probability $P_{\rm BH}$ for the system of two collapsing shells with
$P_1=5\, P_2$ to be a black hole as a function of the total energy of the two shells for $R=30\, \lp$.
\underline{Top right}:
$P_{\rm BH}$ as a function of the mean radius $R$ for $P_1=5\,\mpl$,
$P_2=\mpl$ and $E= 6\,\sqrt{2}\,\mpl\simeq 8.5\,\mpl$.
\underline{Middle left}:
$P_{\rm BH}$ for the system of two collapsing shells with $P_1= P_2$
as a function of the total energy of the two shells for $R=30\, \lp$.
\underline{Middle right}:
$P_{\rm BH}$ as a function of the mean radius $R$ for $P_1=3\,\mpl$,
$P_2=3\,\mpl$ and $E= 6\,\sqrt{2}\,\mpl$. 
\underline{Bottom left}:
$P_{\rm BH}$ for a single collapsing shell as a function of the energy of the shell for a mean radius
$R=30\, \lp$.
\underline{Bottom right}:
$P_{\rm BH}$ for a single shell as a function of the mean radius $R$ for $P=6\,\mpl$ and
$E= 6\,\sqrt{2}\,\mpl$.
}
\label{P_BH_slices_different}
\end{figure*}

To make the results clearer for this case as well, we point our readers to the top plots in
Fig.~\ref{P_BH_slices_different}.
In the top left plot we consider two shells which collide at $R=30\,\lp$.
The probability for the two shells to form a black hole is already more than $50\%$ for a total
energy of $11\,\mpl$.
When using the hoop conjecture to calculate the threshold energy needed to form
a black hole with a radius of $30\,\lp$, the result is $E=15\,\mpl$.
The plot on the top right represents a slice of the three-dimensional plot where the total energy is
constant and equal to $6\,\sqrt{2}\, \mpl\simeq 8.5\, \mpl$.
In this case, the hoop conjecture suggests that the horizon radius should be at $17\, \lp$.
We again notice that the probability for a black hole to form is already larger than $50\,\%$
if the two shells collide at a radius of about $20\,\lp$ and it increases with the decrease of the
mean radius. 
\par
In order to understand how the probability $P_{\rm BH}$ evolves, both as a function of the total energy
and of the mean radius, two more cases were added to the plot in Fig.~\ref{P_BH_slices_different}.
The middle left and right plots represent two collapsing shells of equal momenta.
The bottom plots represent a single collapsing shell.
For consistency, the total energy is the same: for the two-shell scenarios the sum of the two momenta
in each case is the same and it is also equal to the momentum of the single collapsing shell. 
\par
By comparing the three plots on the left we notice that all three cases are fairly similar.
Regardless of whether the same amount of energy is distributed between two shells or it is carried
by a single shell, for the collision taking place at the same mean shell radius the probability $P_{\rm BH}$
increases almost in the same way with the total energy.
For instance, if we evaluate this probability for $E=14\,\mpl$ on all three plots, we obtain
$P_{\rm BH}\simeq 80\%$ in both cases of colliding shells and a slightly larger value when the entire
momentum corresponds to a single shell.
The same argument applies to the three plots on the right: when the total energy is the same,
the probability for a black hole to form varies very similarly with the mean radius of the shells in all three cases.
\subsection{Shells colliding with opposite speeds at finite radius}
\label{S:2opposite}
We now investigate the horizon formation in the case of two shells with radial speeds in opposite directions 
that collide at $R_1=R_2\equiv R>0$.
The wave-functions for the two shells are described by Eq.~\eqref{psi_a}, where one needs to keep
track of the signs of the two momenta.
The HWF can then be calculated in a similar fashion as it was done in the previous section.
One needs to use the upper signs from Eq.~\eqref{fourier_gaussian_approx} for the expanding shell
(with positive $P$) and the lower signs for the collapsing one (with negative $P$). 
The normalised HWF, whose expression we will not write down explicitly due to
its cumbersome mathematical form, is then used to calculate the probability for the
two colliding shells to form a black hole.
\par
\begin{figure*}[h!]
\centering
\includegraphics[width=7cm]{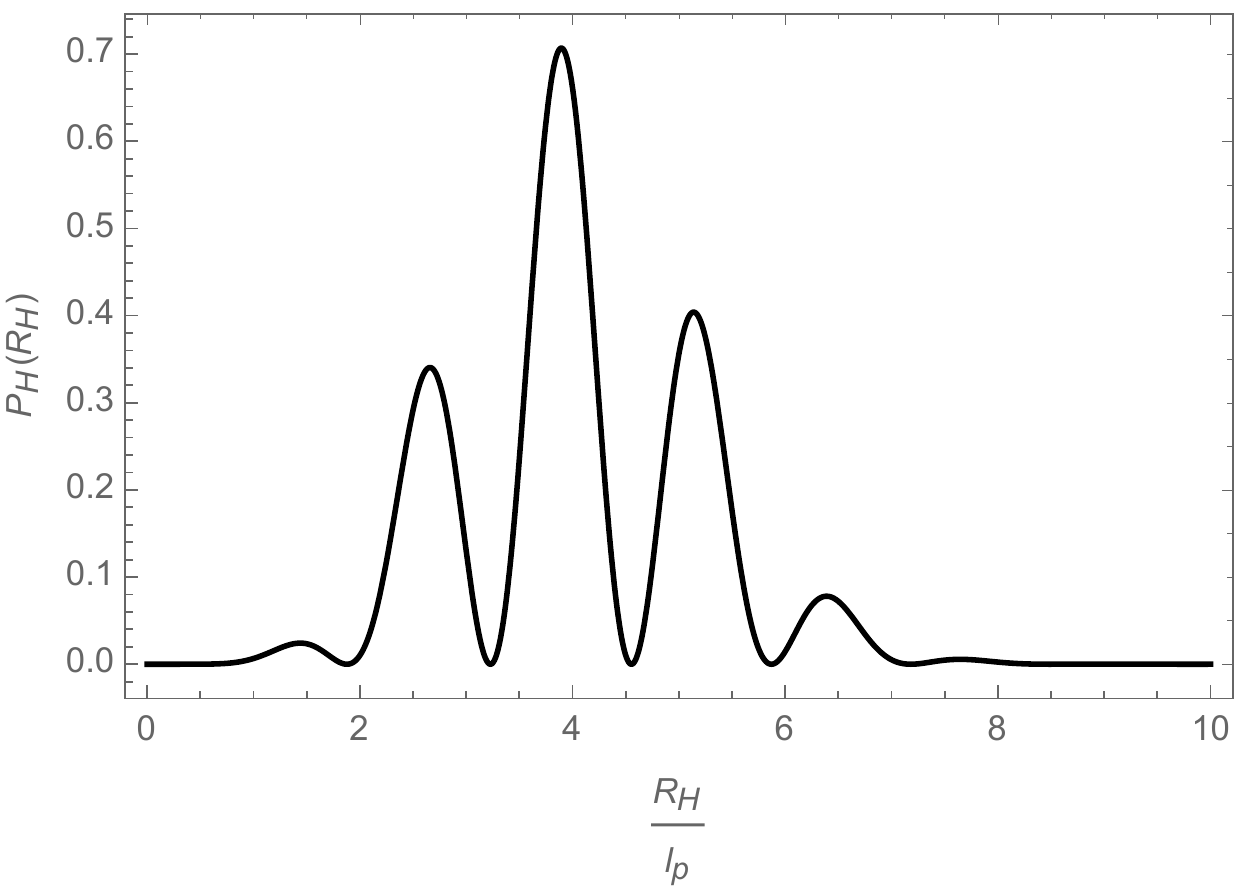}
\includegraphics[width=7cm]{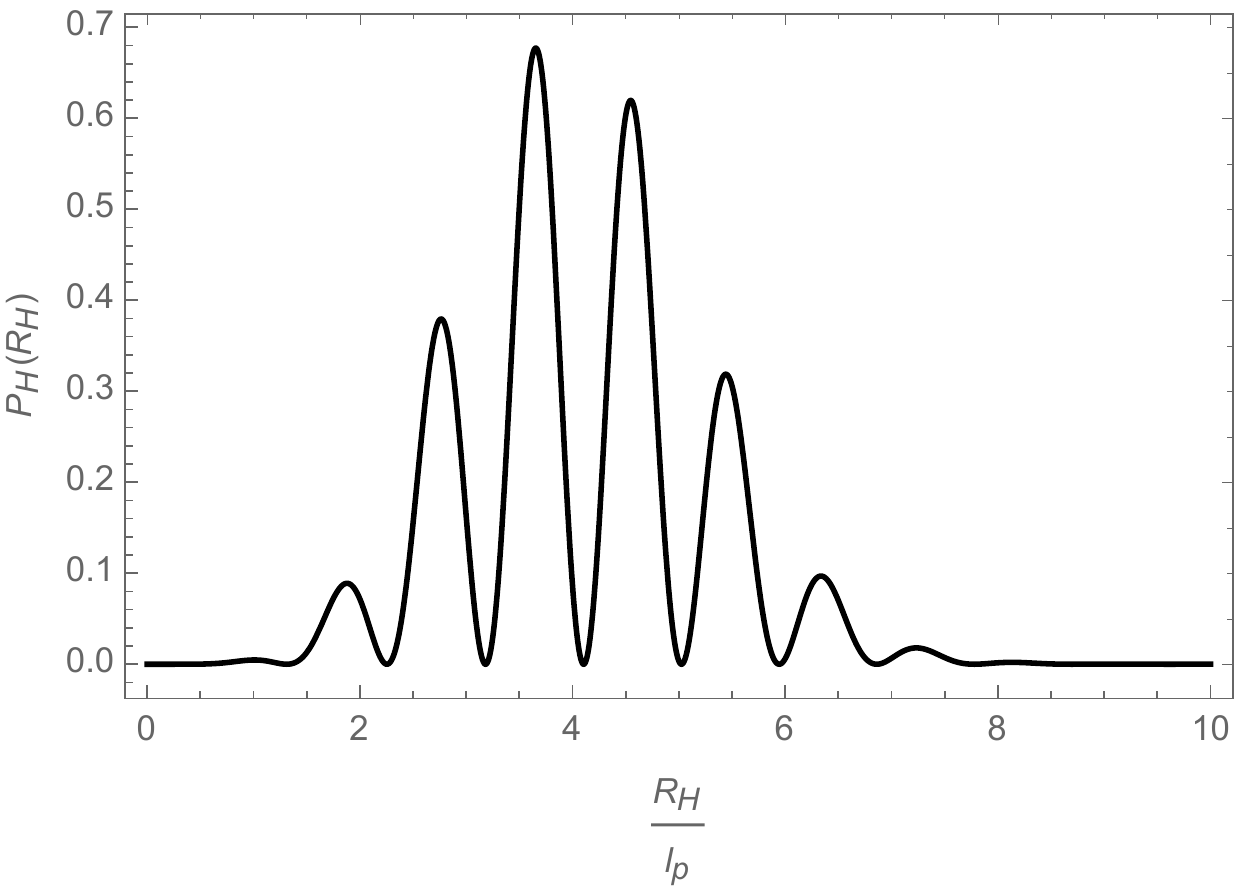}
\\
\includegraphics[width=7cm]{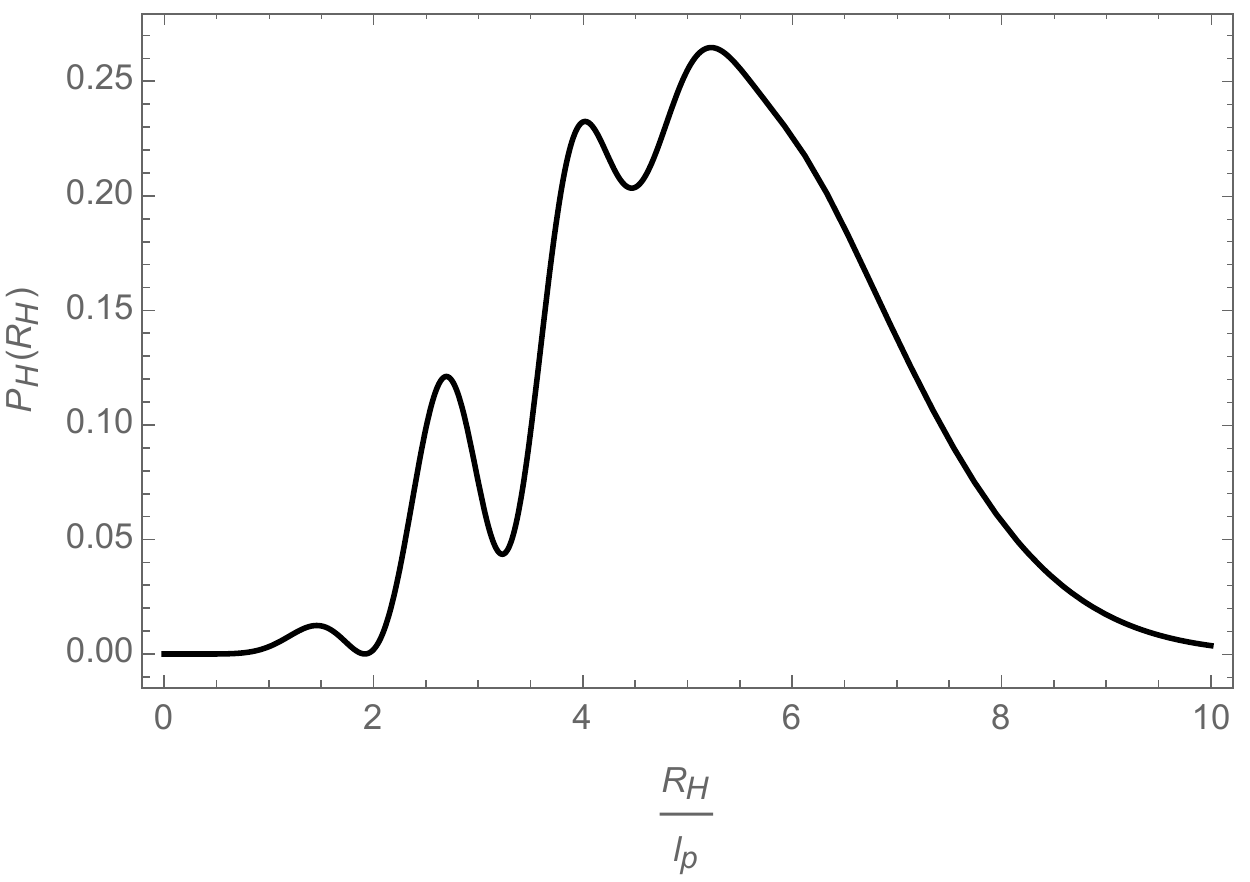}
\includegraphics[width=7cm]{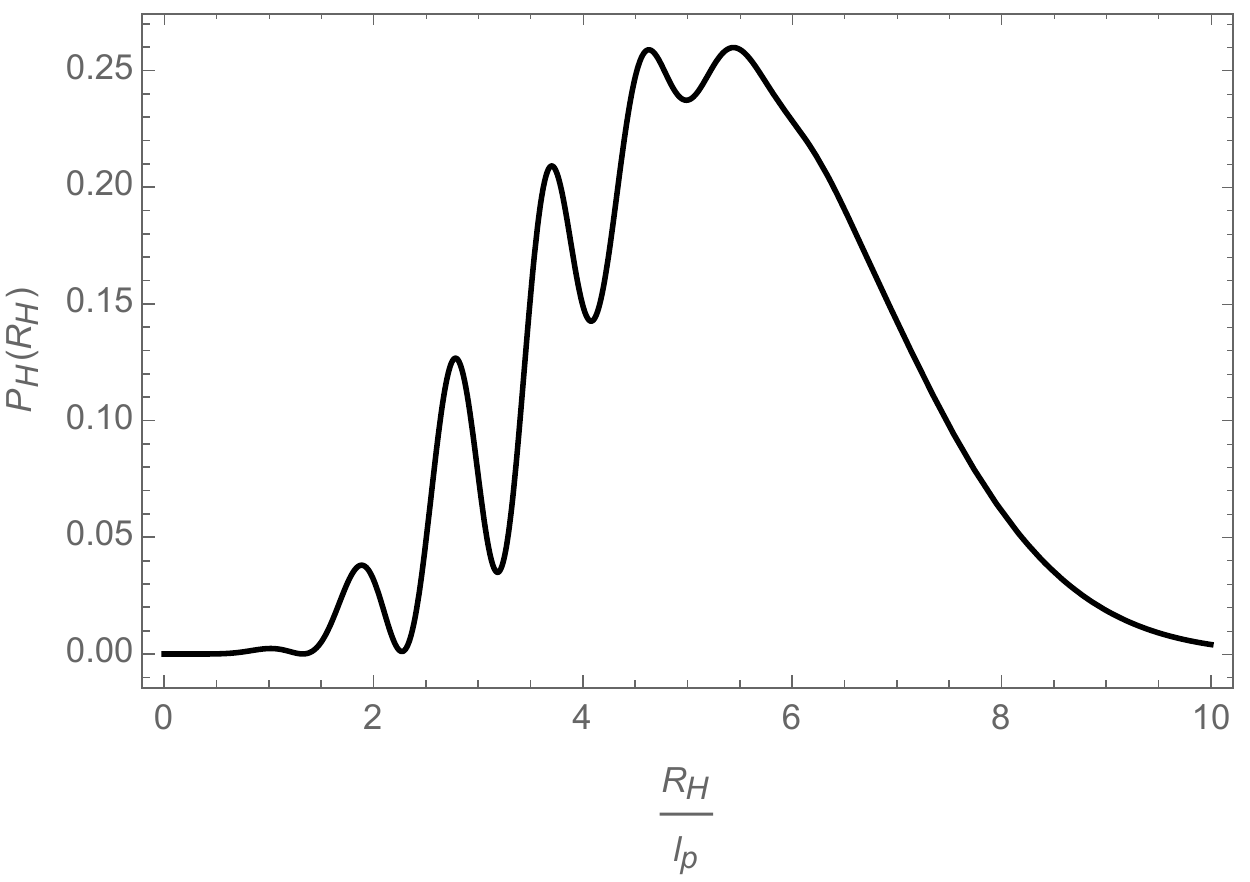}
\caption{Probability density for the horizon to be located on the sphere of radius $r=\rh$ for two shells with
opposite momenta. 
\underline{Top left}: $P_1=P=-P_2$ for a total energy $E=2\,\sqrt{2}\,\mpl\simeq 2.8\,\mpl$ and an average radius
of $R=5\,\lp$. 
\underline{Top right}: $P_1=P=-P_2$ for a total energy $E=2\,\sqrt{2}\,\mpl$ and an average radius of $R=7\,\lp$.
\underline{Bottom left}: $P_1=2\,P$ and $P_2=-P$ for a total energy $E=2\,\sqrt{2}\,\mpl$ and an average radius
of $R=5\,\lp$. 
\underline{Bottom right}: $P_1=2\,P$ and $P_2=-P$ for a total energy $E=2\,\sqrt{2}\,\mpl$ and an average radius
of $R=7\,\lp$.}
\label{PsiPsi_opposite_P0}
\end{figure*}

First we can inspect the probability density ${\mathcal P_{\rm H}}$ in Eq.~\eqref{PhRh} for the horizon to be located
on the sphere of radius $r=\rh$, which is shown in Fig.~\ref{PsiPsi_opposite_P0}.
We have considered four different cases.
The upper plots are obtained for two shells with equal and opposite momenta $P_1 \equiv P=-P_2$
that collide at $R=5\,\lp$, respectively $R=7\,\lp$.
The lower plots represent two shells colliding at the same mean radii as above,
but having different momenta: $P_1 \equiv 2\,P$ and $P_2\equiv -P$.
Unlike the corresponding plots shown in the previous sections, or other cases considered
previously~\cite{Casadio:2013aua, Casadio:2013uga, Casadio:2015rwa, Casadio:2015qaq, Casadio:2017nfg},
this time ${\mathcal P_{\rm H}}$ shows a modulation, with roughly a Gaussian envelope, clearly due to the interference
between the wave-functions of the Gaussian shells in momentum space.
This behaviour is most obvious when the two shells have exactly equal and opposite momenta.
When the momenta of the shells are different, the oscillatory behaviour overlapping the Gaussian profile
becomes more asymmetric, as can be seen from the bottom plots of Fig.~\ref{PsiPsi_opposite_P0}.
It needs to be emphasised that, this behaviour only appears when the two shells collide at a mean radius
larger than zero. 
When comparing the plots on the left to the ones on the right, we notice that ${\mathcal P_{\rm H}}$
oscillates faster with $\rh$ as the mean radius of the collision increases. 
When the momenta are equal in magnitude, the probability density becomes maximum at values of $\rh$
smaller than the ones from Eq.~\eqref{hoop}.
As the difference between the momenta of the shells increases, ${\mathcal P_{\rm H}}$ tends to be
maximum at values closer to $\rh=2\,\lp\,{E}/{\mpl}$, where $E$ is the total energy of the system. 
\par
\begin{figure*}[]
\centering
\includegraphics[width=11cm]{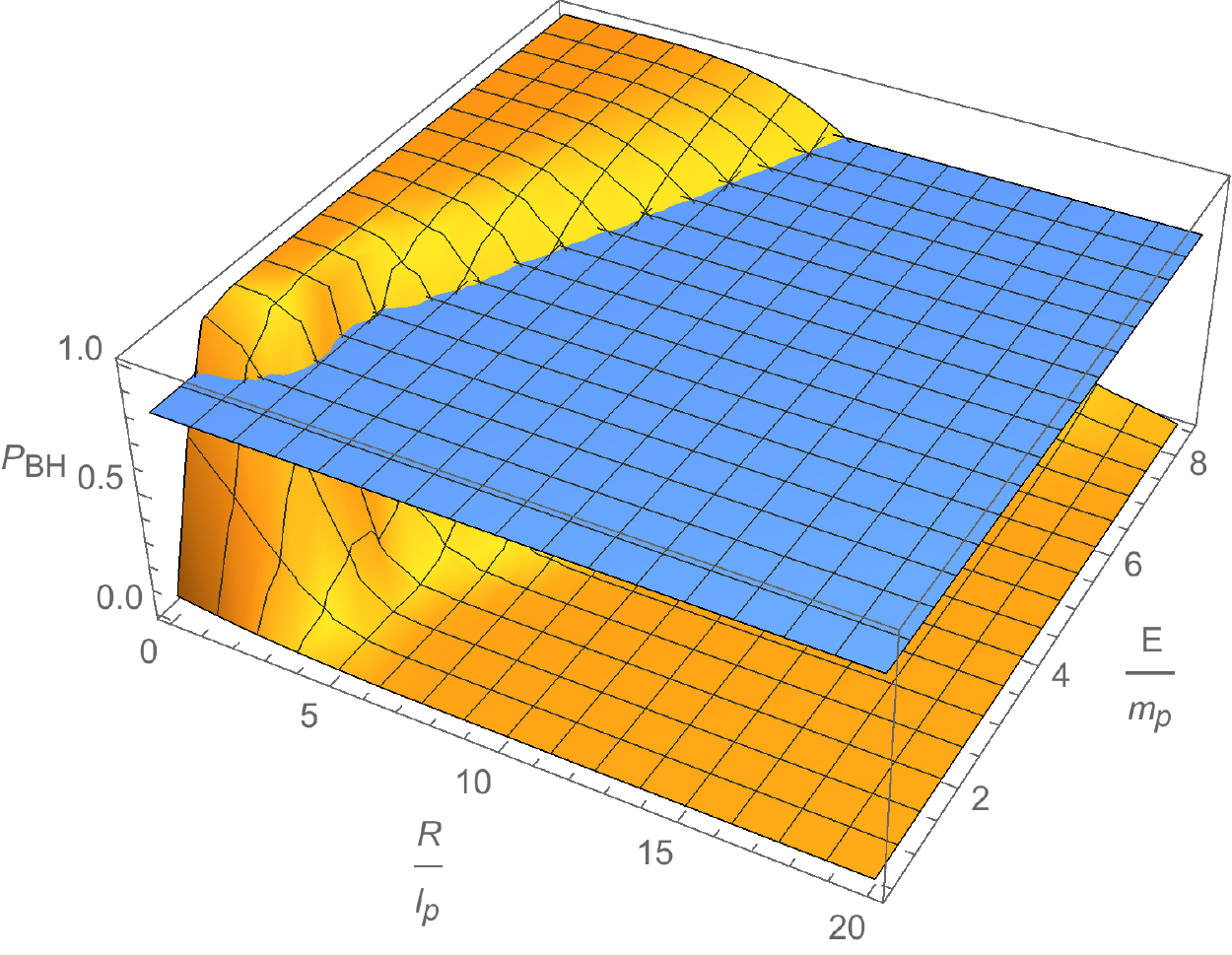}
\caption{Probability for two colliding shells with momenta $P_1=P$ and $P_2=-P$ to form a black hole as a
function of the radius $R$ and the total energy $E$ (in Planck units).
The blue plane delimits the region where the probability $P_{\rm BH}>0.8$.}
\label{P_BH_opposite_P0_different}
\end{figure*}
\begin{figure*}[t]
\centering
\includegraphics[width=7cm]{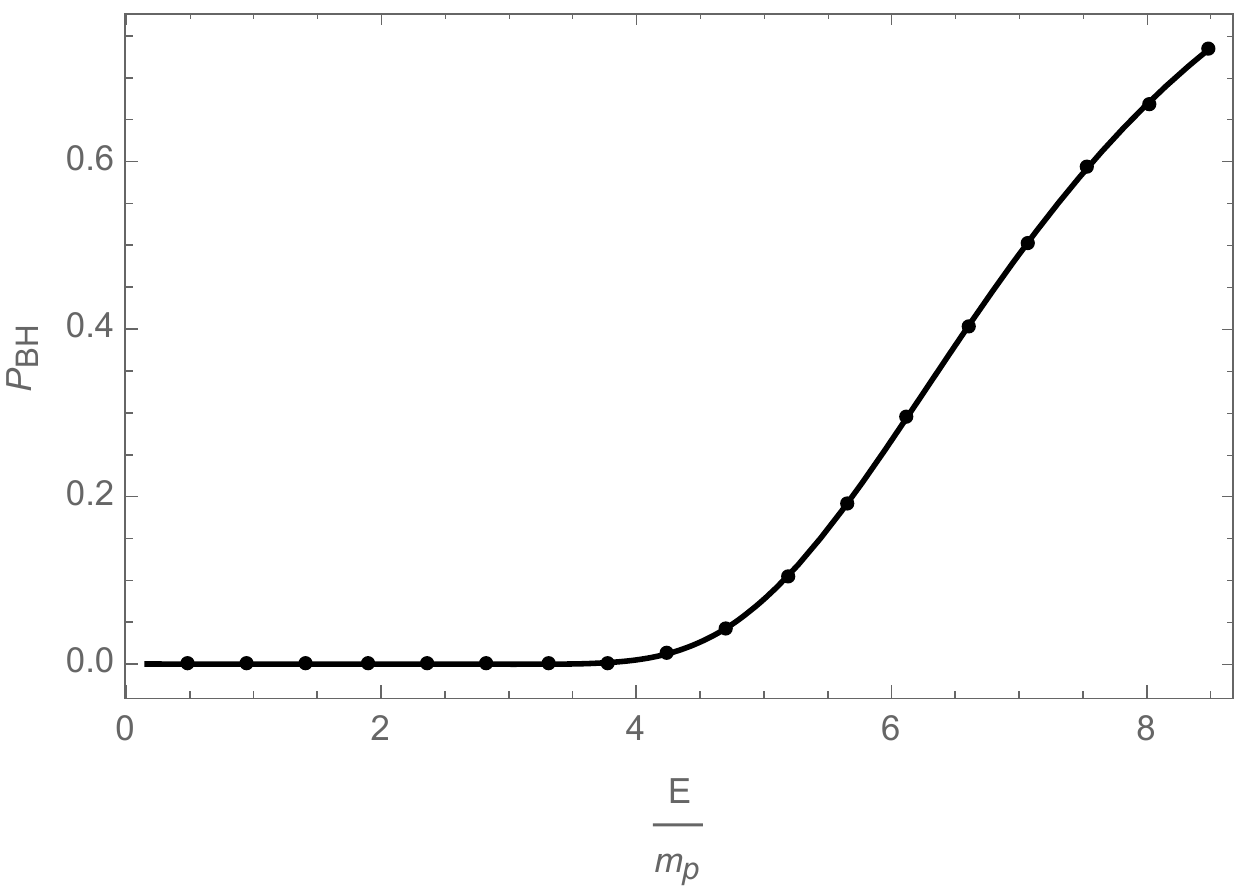}
\includegraphics[width=7cm]{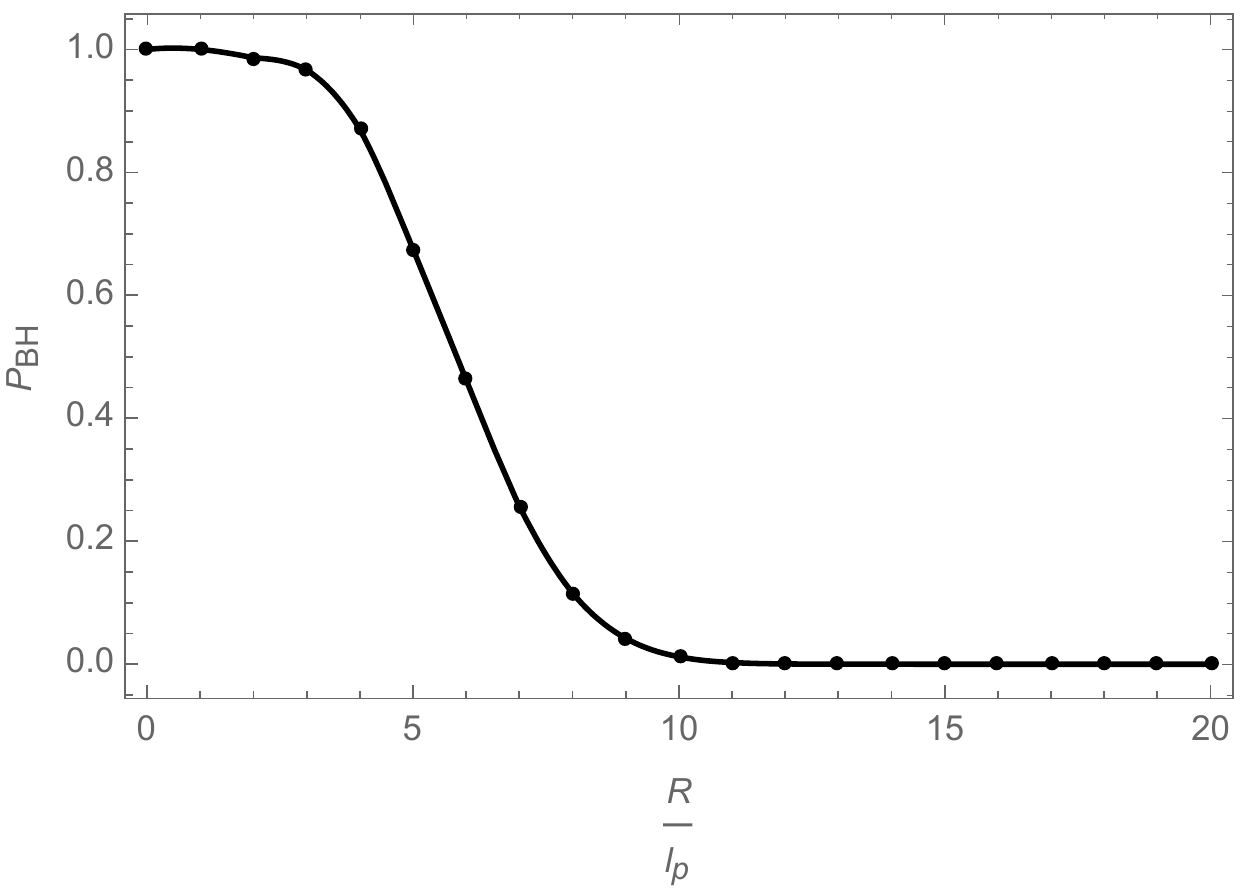}
\\
\includegraphics[width=7cm]{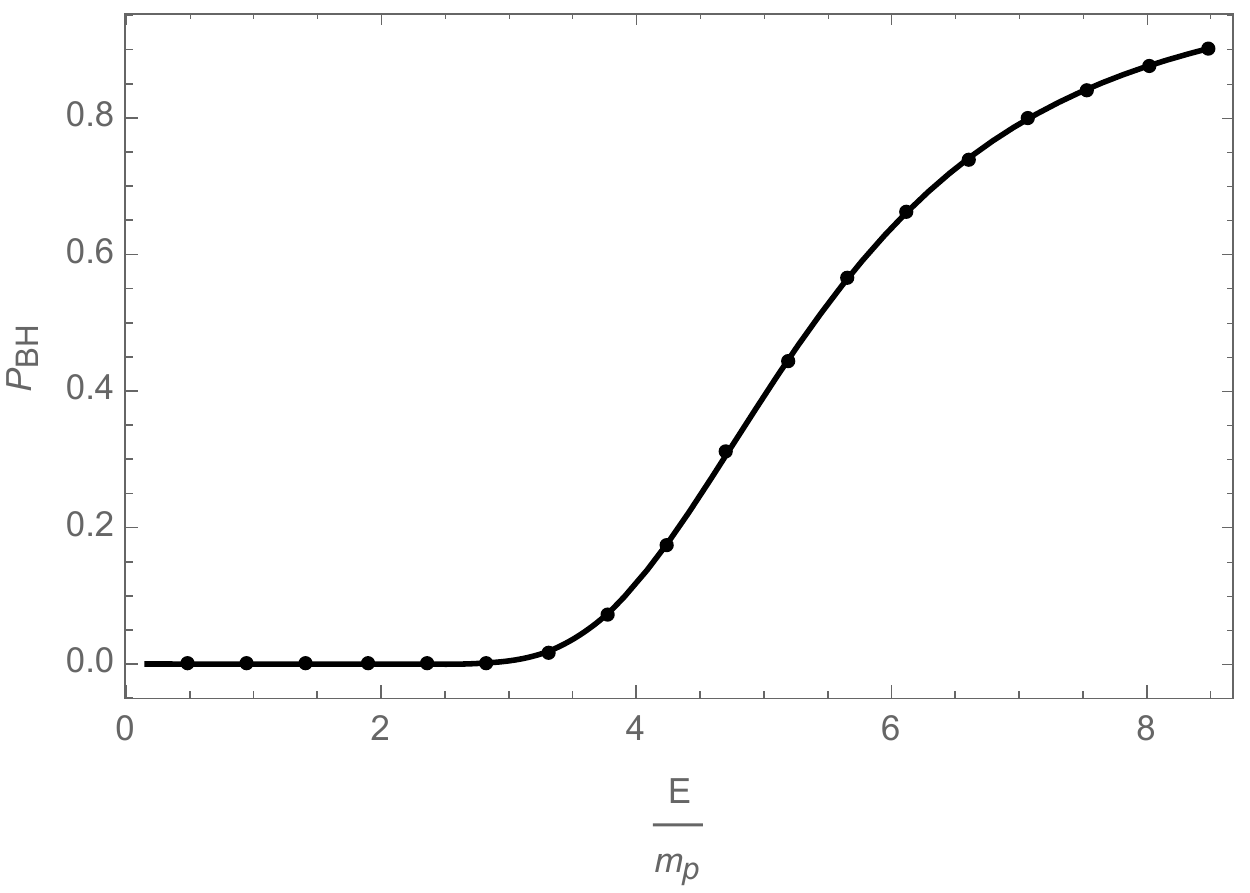}
\includegraphics[width=7cm]{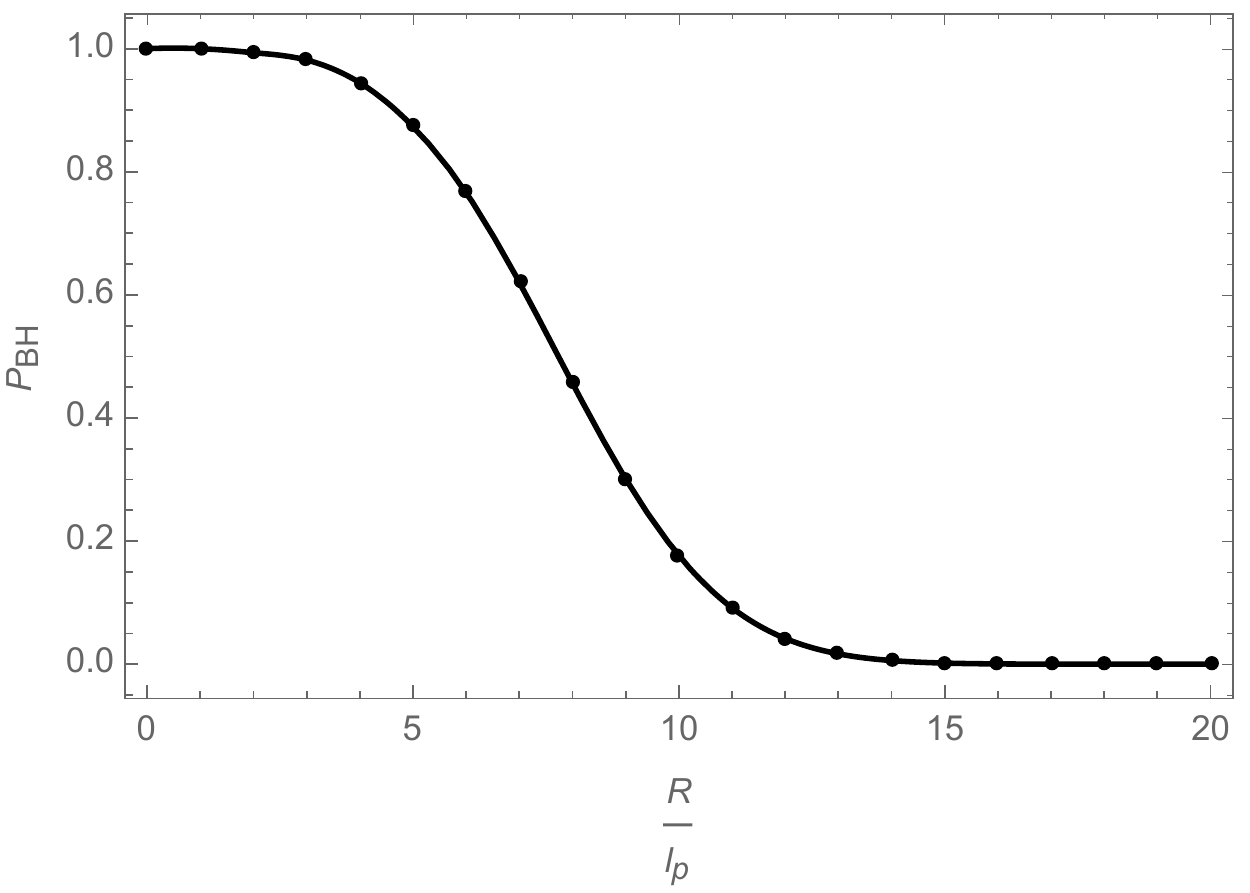}
\caption{\underline{Top left}:
Probability $P_{\rm BH}$ for a system of shells with opposite radial momenta
$P_1= P$ and $P_2=-P$ to be a black hole as a function of the total energy of the two shells for $R=10\, \lp$.
\underline{Top right}:
$P_{\rm BH}$ as a function of the mean radius $R$ for $P_1=P$ and $P_2=-P$ and
$E= 3\,\sqrt{2}\, \mpl\simeq 4.2\,\mpl$.
\underline{Bottom left}:
$P_{\rm BH}$ for a system of shells with opposite radial momenta
$P_1=2\,P$ and $P_2=-P$ to be a black hole as a function of the total energy of the two shells for $R=10\, \lp$.
\underline{Bottom right}:
$P_{\rm BH}$ as a function of the mean radius $R$ for $P_1= 2\,P$ and $P_2=-P$ and  $E= 3\,\sqrt{2}\, \mpl$.
}
\label{P_BH_slices_opposite}
\end{figure*}

A three-dimensional plot of the probability $P_{\rm BH}=P_{\rm BH}(R, 2\,E)$ obtained for equal and opposite
momenta is presented in Fig.~\ref{P_BH_opposite_P0_different}.
When comparing this plot to the one for the two shells having the same velocity in Fig.~\ref{P_BH_same_P0},
we notice that for the same value of the total energy of the system, the probability $P_{\rm BH}$ decreases
approximately twice as fast with the radius  in the case of shells with opposite speeds.
The same conclusion can be drawn when comparing the plots from Fig.~\ref{P_BH_slices_opposite}
with the ones in Fig.~\ref{P_BH_slices}.
Before continuing the analysis, we need to explain that the top plots from Fig.~\ref{P_BH_slices_opposite}
represent two perpendicular slices of the above three-dimensional plot, while the bottom ones are generated
for a similar scenario in which we consider two shells with momenta $P_1 \equiv 2\,P$ and $P_2\equiv -P$.
In all three cases discussed here, including the one from Fig.~\ref{P_BH_slices}, the two shells are assumed
to collide at $R= 10\, \lp$.
By examining them side by side, we observe that the probability $P_{\rm BH}$ increases faster with the energy
of the system when the two shells have parallel radial momenta than when their momenta are opposite.
Moreover, when comparing the two cases from Fig.~\ref{P_BH_slices_opposite}, we see that for the same value
of the total energy the probability is the smallest when the shells have equal momenta, and it increases with the
difference between the momenta of the shells.
This trend was verified to be consistent for larger differences between the momenta of the two shells. 
\par
The probability $P_{\rm BH}$ being the smallest for shells with equal and opposite momenta, and increasing
with the difference between the momenta of the two shells for the same total energy is a particular effect for
systems of shells with opposite momenta.
In Section~\ref{S:2different}, where we compared shells with different radial momenta in the same
direction, we concluded that the probabilities are approximately the same, regardless if the same energy is
carried by a single shell, or distributed between two collapsing shells of equal or different momenta.
As we stated once more, the present case is different, and we attribute this effect to the simple fact that
the wave-functions~\eqref{psi_a} in momentum space are necessarily complex and will give rise to
interference effects, as shown in Fig.~\ref{PsiPsi_opposite_P0}.
A system of nested collapsing shells is indeed the first instance we have encountered which makes this
quantum mechanical feature of the HWF apparent.
\par
Similar conclusions can be drawn from the plots on the right in Fig.~\ref{P_BH_slices_opposite}, which show
that the probability $P_{\rm BH}$ decreases faster with the radius $R$ when the two shells have opposite
radial momenta than when they are changing in the same direction, and that the decrease with the radius is
the fastest when the two shells have equal and opposite radial momenta.
As stated earlier, from the hoop conjecture one estimates that black holes should only form when the total
energy of the two shells overlapping at $R=10\, \lp$ is larger than $5\, \mpl$. 
We notice that in the case of shells with equal and opposite momenta $P_{\rm BH}$ is close to zero for
a total energy of about $5\, \mpl$, while in the other cases this is larger than $40\%$ (it is larger than $80\%$
for shells changing in the same direction). 
Similarly, for the plots on the right, the hoop conjecture suggests a zero probability for the system of shells
to form a black hole if they collide at mean radius values larger than $6 \,\sqrt{2}\,\lp\simeq 8.5\,\lp$.
While for equal and opposite radial momenta this seems to be the case, the plots on the right of
Fig.~\ref{P_BH_slices_opposite} show that the probability is about $40\%$ for shells with different
momenta and Fig.~\ref{P_BH_slices} shows that when the momenta are parallel this probability is up
to $90\%$.
\subsection{Single shell collapsing on a central mass}
\label{S:particularN}
In this section we consider a Gaussian shell collapsing towards a central spherical classical object of mass $m$.
Considering that the gravitational mass of the central object is evaluated by integrating on a flat background
as in Eq.~\eqref{M}, the density profile of the object does not influence the result as long as the probability
for the system to form a black hole is evaluated before or when the collapsing shell of radius $R$ reaches
the edge of the central mass distribution.
The only variables that enter the equation are the total energy contained inside the sphere of radius $R$
and the parameters describing the collapsing shell. 
\par
The configurations considered in Section~\ref{S:2different} can be viewed as a special case in which
a system of $N\gg 1$ shells is divided into a subsystem of $(N-1)$ shells that is described as one macro-shell
(the shell with larger momentum) and a much lighter single shell.
The results presented in this section can also be looked at as a system of $N\gg 1$ shells which is divided into
a subsystem of $(N-1)$ shells that already collapsed and formed the central mass $m$ and an additional lighter
contracting shell.
Both these cases could be of particular interest to investigate the horizon formation caused by the addition
of a small amount of energy to a macroscopic system very near the threshold of forming a black hole.
\par
\begin{figure*}[]
\centering
\includegraphics[width=11cm]{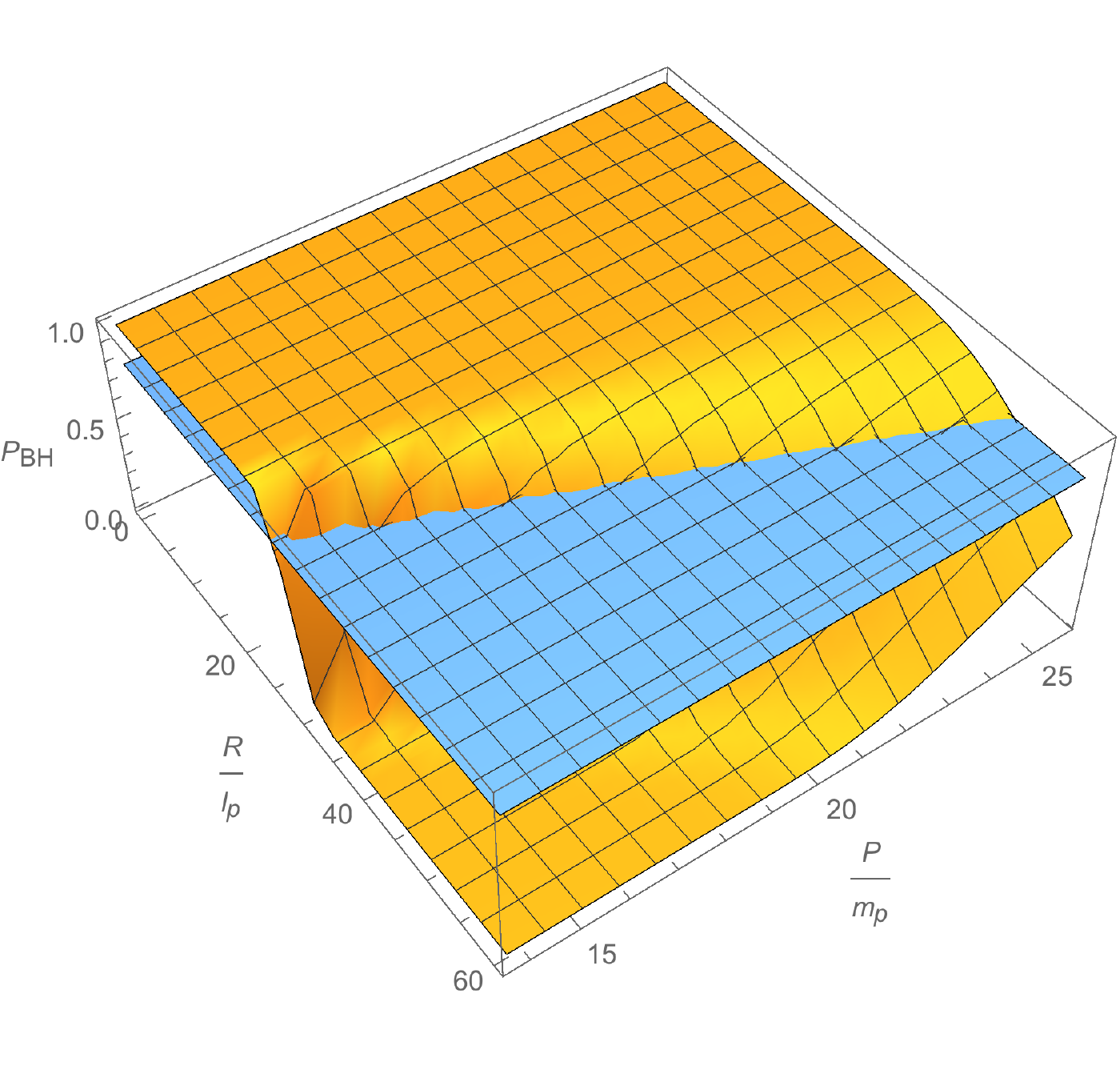}
\caption{Probability for a shell collapsing towards a central mass $m$ to form a black hole as
a function of the radius of the central mass $R$ and the total energy of the system $E$
(in Planck units).
The blue plane delimits the region where the probability $P_{\rm BH}>0.8$.}
\label{PBH_shell_mass}
\end{figure*}

Due to the analogy between this case and the one from Section~\ref{S:2different},
we consider values for the central mass and shell momentum that will make comparisons between the two
cases easy to read.
Fig.~\ref{PBH_shell_mass} shows the three dimensional plot of the probability
$P_{\rm BH}=P_{\rm BH}(R, m+E)=P_{\rm BH}(R, m+\sqrt{2}\,|P|)$ for the system to form a black hole
for a value of the central mass of $m= 13\,\mpl$ and the energy of the collapsing shell varying from
$0.5\,\sqrt{2}\,\mpl$ to $9\,\sqrt{2}\,\mpl$.
The plot shows that, when the collapsing shell has very little energy, the probability for the object to become
a black hole is zero except when the radius of the central object is about $26\,\lp$, which is the Schwarzschild
radius that corresponds to a mass of $13\,\mpl$.
This is, of course, the expected behaviour for this limit.
Regardless of the momentum of the infalling shell, there is an object that has a radius of roughly the size
of its horizon radius. 
\par
For the same value of the central mass, for larger values of the momentum of the infalling shell
(reflected in the larger value of the total energy of the system), the radius at which the probability
for a black hole to form approaches one also becomes larger. 
We would like to make a remark here.
As detailed earlier, for this plot we supposed that the central mass occupies the whole volume up to the
radius $R$ at which the collision occurs.
In principle, considering that the radius of the central object does not play a role in calculating this
probability (we only have to make sure that it is smaller or at most equal to the radius of the shell),
the central mass could be contained within a smaller volume than the one enclosed by the collapsing
shell and the probability $P_{\rm BH}$ would increase in the same way with the momentum of the
shell and its radius $R$. 
\section{Conclusions and outlook}
\label{S:conc}
Thin shells of matter in general relativity provide a very simple, yet useful way of modelling processes that might
occur inside compact astrophysical objects which collapse and form black
holes~\cite{Kijowski:2005qm,Cardoso:2016wcr}.
In this work, we have started to investigate the quantum dynamics of shells by means of the HQM applied to two
main scenarios:
a) two ultra-relativistic shells colliding into each other and b) one shell collapsing towards a larger central source. 
\par
The first scenario regards collisions between two Gaussian shells that reach the same mean radius
and was studied in Sections~\ref{S:2equal}-\ref{S:2opposite}. 
Two types of situations can be distinguished:
shells which both shrink, so that their radii changes in the same direction, and shells whose radii 
change in opposite directions.
For shells shrinking in the same direction the results show that, independently of the way in which the energy
is distributed between the two shells or if it is carried by a single shell, for the same mean shell radius the
probability for black holes to form increases approximately in the same way with the total energy.
Moreover, the probability $P_{\rm BH}$ increases gradually from zero to one as the energy of the system
increases, and is still larger than zero at energies below the minimum value estimated using the hoop conjecture
at face value.
For instance, we see from Fig.~\ref{P_BH_slices_different} that for energies smaller than $15\,\mpl$
(the minimum energy at which black holes with radii of $30\,\lp$ should form in a classical scenario)
this probability already increased to about $80\%$. 
\par
The above findings are consistent with what was obtained previously for other cases investigated
using the HQM. 
The case that stands out is the one of shells that collide with radii changing in the opposite directions.
First we need to draw the reader's attention to the interference effects that lead to the modulation
of the probability density for the horizon to be located on the sphere of radius $r=\rh$.
This is the first instance in which this behaviour was observed. 
When comparing this case with the previous ones, considering the same mean radius at the instant
of the collision, the probability for black holes to form increases slower with the energy.
When considering the same total energy of the system, the probability $P_{\rm BH}$ drops
to values below $80\%$ at radii that are almost half of the radii at which this happens in the other
cases.  
Not only this, but the way the energy (roughly equal to the momentum in the ultra-relativistic limit)
is distributed between the two shells affects the probability
for black holes to form as well, which also differs from what was found earlier for simpler 
configurations.
All other parameters such as mean radii and total energy of the system being equal, $P_{\rm BH}$
is the smallest for shells with equal and opposite momenta and it increases as the difference between
the momenta of the two shells gets larger.
We attribute this effect to the interference between the wave-functions of the two Gaussian
shells in momentum space, and this is an effect that could not be explained in a classical scenario. 
\par
The case b) presented in Section~\ref{S:particularN} can be looked at as a separate category from the rest,
because it does not involve a collision between two shells of matter.
In this case, we considered a spherical Gaussian shell that is collapsing towards a central mass.
The central mass can be a classical spherical object of mass $m$ or even a shell, or system of shells,
of radii smaller than the collapsing shell having a total energy $E\sim m$.
The smallest horizon radius possible, when the shell carries a negligible amount of energy, is the one
corresponding to the central mass, and a black hole can form provided the mass is located within its
Schwarzschild radius.
Otherwise, the size of the horizon increases with the total energy of the system.
However, the probability function for a horizon to form (or the system to be a black hole) is not a step function
as the classical hoop conjecture would suggest.
The probability $P_{\rm BH}$ is again a smooth function that increases from small values of the total
energy, when the mean radius of the shell is larger than the classical value of the horizon corresponding to the
total energy of the system, and approaches one when the mean radius of the shell is smaller than said
classical value. 
\par
Both scenarios a) and b) above could be useful for our understanding at the quantum level of the
formation of a (apparent) horizon inside a collapsing astrophysical body.
Of course, any realistic modelling of such an event requires heavy numerical calculations
already at the classical level of the Einstein equations sourced by a fluid with given equation of
state.
We believe it would be very interesting to try and analyse whether the HQM description for the
horizon formation could be incorporated into such numerical codes and whether the quantum
nature of matter could lead to any significant departures from the purely classical expectations. 
\section*{Acknowledgments}
R.C.~is partially supported by the INFN grant FLAG and also carries his work
in the framework of activities of the National Group of Mathematical Physics
(GNFM, INdAM). O.M.~is supported by the grant Laplas V of the Romanian National Authority for Scientific Research. 
The work has been carried out in the framework of the COST action {\em Cantata\/}. 
\end{document}